\documentclass[12pt,a4paper]{article}
\pdfoutput=1
\usepackage{jheppub}
\usepackage{epstopdf}
\usepackage{graphicx}
\usepackage{epsfig}
\usepackage{dcolumn}  
\usepackage{bm}    
\usepackage{amssymb} 
\usepackage{amsmath,bm}
\usepackage{amsfonts}    
\usepackage{slashed}  
 \usepackage{relsize}
\usepackage[mathscr]{euscript}
\usepackage{epsfig}
\title{R\'{e}nyi entropies of free bosons on the torus and holography }

\author{Shouvik Datta and} 
\author{Justin R. David}

\affiliation{ Centre for High Energy Physics,
Indian Institute of Science,\\ C.V. Raman Avenue, Bangalore 560012, India}

\emailAdd{shouvik, justin@cts.iisc.ernet.in}

\abstract{We  analytically evaluate  the R\'{e}nyi entropies for the two dimensional  free 
boson CFT. The CFT is considered to be compactified on a 
circle and at finite temperature.  The R\'{e}nyi entropies $S_n$ are evaluated for a single interval 
using the two point function of bosonic  twist fields on a torus.  
For the case of the compact boson, the sum over the classical saddle points results in the 
 Riemann-Siegel theta function associated with the $A_{n-1}$ lattice. 
  We then study the R\'{e}nyi entropies in the  decompactification
regime. We  show that in the limit when the size of the interval becomes 
the size of the spatial circle, the entanglement entropy reduces to the thermal 
entropy of free bosons on a circle.  We then set up a systematic high temperature 
expansion of the R\'{e}nyi entropies  and evaluate 
the finite size corrections for free bosons. 
Finally we compare these finite size corrections both for the free boson CFT and the 
free fermion CFT with the one-loop corrections obtained from 
bulk three dimensional  handlebody 
 spacetimes which have   higher genus  Riemann surfaces as its boundary.   
 One-loop corrections in these geometries are entirely determined by quantum numbers
 of the excitations present in the bulk.  This implies that 
 the  leading finite size corrections contributions from 
 one-loop determinants of the Chern-Simons gauge field and the Dirac field in the dual 
geometry   should reproduce that of  the free boson and the free fermion  CFT respectively. 
By evaluating these corrections both in the bulk and in the CFT explicitly we show that 
this expectation is indeed true. 
}

\begin{document}
 \maketitle

\def\nn{\nonumber}
\def\pd{\partial}
\def\Re{R\'{e}nyi }
\def\l1{{\text{1-loop}}}
\def\uy{u_y}
\def\ur{u_R}
\def\o{\mathcal{O}}
\def\Cl{{{cl}}}
\def\bz{{\bar{z}}}
\def\by{{\bar{y}}}
\def\bX{\bar{X}}
\def\im{{\text{Im}}}
\def\re{{\text{Re}}}
\def\cn{{\text{cn}}}
\def\sn{{\text{sn}}}
\def\dn{{\text{dn}}}
\def\K{\mathbf{K}}
\def\n1{\Bigg|_{n=1}}
\def\fin{{\text{finite}}}
\def\R{{\mathscr{R}}}
\def\one{{(1)}}
\def\zero{{(0)}}

\section{Introduction}

Entanglement entropy is emerging as  an important observable to characterize behaviour
of quantum field theories and many body theories. 
In condensed matter physics entanglement entropy is used as an order parameter 
to characterize quantum phase transitions \cite{Osterloh}\footnote{See \cite{Amico:2007ag} for a complete list of references 
for the application of entanglement entropy to many body physics.}. 
It is also useful as a measure of thermalization in non-equilibrium statistical 
mechanics \cite{Calabrese:2005in}. 
Entanglement entropy is defined as follows 
\cite{Bombelli:1986rw,Srednicki:1993im,Holzhey:1994we}: 
First partition the complete set of observables of the quantum systems into two 
disjoint Hilbert spaces $A$ and its complement $B$. 
Define the reduced density matrix $\rho_A$ by tracing over the observables in $B$. 
Then entanglement entropy   between the observables in the 
two Hilbert spaces is defined by 
\begin{equation}
 S_E = {\rm Tr} ( \rho_A \ln \rho_A) . 
\end{equation}
In quantum field theories usually the Hilbert spaces $A$ refers to the degrees of freedom in 
a spatial region $A$. 
Evaluation of entanglement entropy is  usually done by 
computing the R\'{e}nyi entropies which are defined as
\begin{equation}
 S_n = -\frac{1}{n-1} \ln {\rm Tr} ( \rho_A^n) 
\end{equation}
Then entanglement entropy is then derived as the limit $S_E = \lim_{n\rightarrow 1} S_n$. 

Though entanglement entropy has proved to be an useful observable, it is  difficult to 
compute even in the case of free theories.  
In $1+1$ dimensional conformal field theories if $A$ denotes a single interval
 of length $L$, the evaluation  of R\'{e}nyi entropies can be done exactly using the
replica trick \cite{Holzhey:1994we,Calabrese:2004eu,Calabrese:2009qy,Casini:2009sr}. 
The R\'{e}nyi entropies $S_n$  are determined by evaluating the partition 
function of the CFT on a  $n$ Riemann sheets joined consecutively and cyclically along the 
interval $A$. In general these Riemann sheets form a higher genus surface. 
 This can then be shown to be equivalent to evaluating the two 
point function of the `twist' operator of dimension 
\begin{equation}
 \Delta_n  = \bar \Delta_n = \frac{c}{24} \left( n - \frac{1}{n} \right) , 
\end{equation}
inserted at the end points of the interval.  Here $c$ refers to the central charge of the 
CFT.  We therefore obtain 
\begin{eqnarray}
 \exp( -(n-1) S_n (A) )   &=& \langle {\cal T}_n( L)  \bar {\cal T}_n(0) \rangle, \\ \nonumber
&=& \left( \frac{a}{L} \right)^{4\Delta_n}  
\end{eqnarray}
where ${\cal T}_n $ is the twist operator and  $a$  the cut-off. We have normalized the two point functions to unity.  Change in this normalization just leads to a non-universal constant. 
Thus the R\'{e}nyi entropies for a single interval are given by 
\begin{equation} \label{sinter}
 S_n(A)  = \frac{c (n+1) }{6n }  \ln \frac{L}{a}
\end{equation}
Though these are evaluated for an integer $n$, the resulting expression is assumed to be analytic
in $n$ and the entanglement entropy is obtained in the $n\rightarrow 1$ limit. 
Note that the only information  R\'{e}nyi entropies  of a single interval 
carry about the conformal field theory is its central charge. 

There are two possible methods to obtain more information regarding the details of the  CFT using 
R\'{e}nyi entropies. One is to generalize the subsystem $A$ to multiple 
intervals \cite{Calabrese:2009ez,Calabrese:2010he,Headrick:2010zt,  Coser:2013qda}. 
The other is to consider the CFT on a circle of radius $R$ and  at finite temperature $T$, that 
is  the CFT is on a torus. 
R\'{e}nyi entropies for CFT's on tori contain finite size corrections which are 
suppressed by $\mathcal{O}(e^{-\pi RT})$ in the high temperature expansion compared to the 
result in    (\ref{sinter}).  These depend on the details of the CFT. 
In this regard, the only known analytical result for R\'{e}nyi entropies   is the case of 
free fermions on the torus \cite{Azeyanagi:2007bj,Herzog:2013py}. 
In this paper we generalize this to the case of a free complex  boson on 
the torus\footnote{Numerical results for the entanglement entropy of the massive scalar in 1+1 dimensions were obtained in \cite{Herzog:2012bw}.}. 
R\'{e}nyi entropies  corresponding to a 
single interval is evaluated by using the two point function of the 
bosonic twist operators on the torus. 
These two point functions can be evaluated by the `stress-tensor method' \cite{Dixon:1986qv,Atick:1987kd,Saleur:1987tn}  (see also \cite{Dijkgraaf:1987vp}). 
As far as we are aware this is the first instance of an application of 
the bosonic twist correlators on the torus. 
The R\'{e}nyi entropies, $S_n$  consists of two contributions one due to the 
quantum part of the two point function and the other due to the 
various saddle points of the classical action.  
For a boson compactified on a spacetime square torus of size $\R$  and 
for a rectangular worldsheet torus we show that the contribution of the classical
saddle points to the two point function reduces to  the Riemann-Seigel 
theta function associated with the $A_{n-1}$ lattice.
For the interested reader the result is given in (\ref{renyi-full}). 
The compact free boson CFT  describes  a class of  
$1+1$ dimensional condensed matter systems called
Luttinger liquids. Examples of these are 
Heisenberg spin chains and 1-dimensional Bose gases with repulsive interactions. 
At the conformal point they can be described by compact bosons but with different 
radius $\R$. Therefore our result is relevant for all these systems. 

We then study the de-compactification regime in detail. The result for the 
R\'{e}nyi entropies  in this regime is given in (\ref{renyi-1}). 
  We first show that  when the size of the interval $L$ approaches the size of the circle $R$, the 
entanglement entropy  $S_1$  reduces to the thermal entropy of 
a complex boson on a circle.  This  serves a  simple check of our result.
 We then set up a systematic high temperature expansion of the 
R\'{e}nyi entropies  and obtain the leading finite size corrections.

The second goal of this paper is to compare the finite size corrections 
for both the free boson theory and for the free fermion theory 
 to that obtained from holography. 
Though entanglement entropy is a difficult quantity to evaluate
in field theories, there is a very simple proposal by Ryu and Takanayagi to evaluate it
in strongly coupled field theories which admit a gravitational dual
 \cite{Ryu:2006bv,Ryu:2006ef}. 
Recently for the case of conformal field theories in 2 dimensions which admit a 
gravitational dual, the Ryu-Takanayagi   proposal has been derived from 
the conventional rules of $AdS/CFT$
\cite{Faulkner:2013yia}. This derivation has been extended to 
higher dimensions in \cite{Lewkowycz:2013nqa}.  
Recall that R\'{e}nyi entropies are determined by evaluating the partition function
of the CFT on a $n$-sheeted Riemann surface. 
The $n$-sheets are joined consecutively and cyclically along 
the interval $A$. 
From  the rules of 
$AdS/CFT$ we obtain
\begin{eqnarray}\label{adscft}
 Z_n(CFT)  &=& \exp \left( -(n-1) S_n(A) (CFT)  \right) = \int {\cal D }[g] e^{-{\cal S}[g] }, \\ \nonumber
 &=& \exp\Big( c \, {\cal S}_{\text{classical}}  + {\cal S}_{\text{1-loop}}  + \mathcal{O}\left(\tfrac{1}{c}\right)  \Big) 
\end{eqnarray}
where $g$ refers to all fields which occur in the bulk action. The path integral over
$g$ is done over bulk geometries whose boundaries reduce to the $n$-sheeted 
Riemann surface which on which the CFT lives.  ${\cal S}_{\text{classical}}$ is obtained 
by evaluating the classical action on these bulk geometries. $c$ is the central charge 
which is related to the  inverse of the  three dimensional 
Newton's constant. ${\cal S}_{\text{1-loop}}$ refers to the one-loop contribution on 
performing the path integral.  Explicit bulk duals of CFT's 
 living on the replica geometry 
which in general is a higher genus Riemann surface were constructed in \cite{Faulkner:2013yia}.
These bulk duals are handlebody geometries which can be constructed by 
realizing the higher genus Riemann surface at the boundary using 
Schottky uniformization \cite{Krasnov:2000zq}.  The partition function of the CFT is 
 evaluated  holographically by computing the path integral in (\ref{adscft}).  
 Using the equality in (\ref{adscft}) the leading contribution of the R\'{e}nyi entropies
 at large $c$ can be obtained by evaluating ${\cal S}_{\text{classical}}$.
  In the $n\rightarrow 1$ limit it has been   shown to reduce to the
 Ryu-Takanayagi  prescription.

 Using this approach a method  to obtain  one-loop corrections to the  R\'{e}nyi entropies
 as well as the 
 Ryu-Takanayagi formula has been proposed in 
 \cite{Barrella:2013wja}\footnote{An alternate method to 
 evaluate one loop corrections has been proposed by 
 \cite{Faulkner:2013ana} for higher dimensions. This has been verified recently for the case of 
 3 dimensional CFT's in \cite{Swingle:2013rda}.}. 
 The proposal has been verified upto the 6th order in the short interval expansion of the Renyi entropy of two disjoint intervals on a plane in \cite{Chen:2013kpa}.
  This proposal also predicts finite size corrections to the R\'{e}nyi entropies of a single
 interval on a torus. Determining the 1-loop corrections ${\cal S}_{\text{1-loop}}$ involves 
 evaluating one-loop determinants of all the fields present in the bulk handlebody 
  geometries. One loop determinants in these handlebody geometries have been 
 evaluated in \cite{Giombi:2008vd} following a conjecture by 
 \cite{Yin:2007gv}.  The one-loop determinants
  are sensitive to only the quantum numbers of the spectrum of the  
 fields in the bulk. That is,  given the spin and the mass  of the bulk field, the 
 one-loop determinant is completely specified. One-loop determinants are in-sensitive 
 to interactions. 
 In fact the determinant organizes itself into  
  characters of the Virasoro algebra as expected from 
  symmetry considerations 
  Further more the 1-loop contribution is independent of the 
 central charge of the theory. 
 In fact when the bulk geometry is thermal $AdS_3$ which is dual to the CFT on a genus one surface,
 it has been seen that the 1-loop determinant of a single field in the bulk organizes it self into
 Virasoro blocks and is equal to the partition function of the dual field in the CFT 
  \cite{Maloney:2007ud,Giombi:2008vd,David:2009xg,Gaberdiel:2010ar}. 
 Note that the dual operator  is also entirely
 determined by the quantum numbers of the corresponding field in the bulk. 
 Thus we expect the  contribution to the 
 finite size corrections  of  the R\'{e}nyi entropies from the 1-loop determinant  
 of a single $U(1)$  Chern-Simons field should capture the  leading finite 
 size corrections of the R\'{e}nyi entropies of single  boson.
 The  Chern-Simons field in the bulk is  dual to  currents 
 with conformal 
 dimensions $(1, 0)$ and $(0,1)$  which are the physical degrees of freedom of a  free boson CFT.
 Similarly we expect the  leading  contribution to the finite size corrections 
 of the R\'{e}nyi entropies from the 1-loop determinant of 
 a single Dirac field to
   be captured by the  free fermions with  conformal dimensions  $(1/2, 0)$ and $(0,1/2)$. 
   It is important to   note that 
 the classical action ${\cal S}_{classical}$ also contains finite size 
 corrections to R\'{e}nyi entropies. We will see that these finite size corrections 
 are sub-leading  in $e^{-\pi RT}$ when compared to that from the one loop determinants for both 
 the Chern-Simons field as well as the Dirac field in the bulk.  These finite size corrections also 
 do not contribute to the entanglement entropy as they vanish in the $n\rightarrow 1$
 limit\footnote{The finite size contributions to the R\'{e}nyi entropies 
 from the 1-loop determinant of 
 the graviton in the bulk begins at  the same order in $e^{-\pi R T }$ as the finite
 size contributions from ${\cal S}_{\text{classical}}$. Therefore in this case the leading 
 finite size corrections cannot be obtained just from  ${\cal S}_{\text{1-loop}}$.}.
  
   Motivated by these arguments we evaluate 
   the leading finite size corrections obtained 
  by considering the one loop determinant of the Chern-Simons 
  gauge field in the handlebody geometries. We show that these corrections  agree precisely with the 
  leading finite size corrections for the case of free bosons on the torus. 
  We also compare the finite size corrections obtained 
  by considering the one loop determinant of the  Dirac field on the 
  handle body geometries and obtain precise agreement with the 
  leading finite size corrections for the case of free fermions on the 
  torus.  We consider these agreements as a non-trivial check of the method proposed in 
  \cite{Barrella:2013wja}.

The organization of the paper is as follows. In the next section we derive the 
R\'{e}nyi entropies of a free  compact  boson for a single interval on a torus. 
We detail the necessary methods from the theory of $\mathbb{Z}_n$ orbifolds for this purpose. 
In section 3 we discuss the decompactification regime in detail and 
show that in the limit when the size of the interval becomes the size of the 
CFT spatial circle the entanglement entropy reduces to the thermal entropy of bosons
on a circle. 
In section 4 we write down the leading  finite size corrections to 
R\'{e}nyi entropies for 
free bosons on the torus and for the free fermions in the high temperature 
expansion. The result for the 
free fermions has been derived  earlier in \cite{Azeyanagi:2007bj} 
 and  we review  it here  for completeness. 
In section 5 we evaluate finite size corrections to R\'{e}nyi entropies  in the bulk handlebody 
geometries using the method proposed in \cite{Barrella:2013wja}.
We show that  finite size corrections  obtained 
by considering one-loop determinants of the Chern-Simons gauge field 
and the Dirac field agree precisely with the leading finite corrections obtained 
for the free boson and free fermion CFT. Finally, section 6 contains our conclusions.
Appendix A contains details of conventions we use for Jacobi-theta functions. 
Appendix B derives various identities for integrals of cut differentials used in 
the main text. Appendix C contains the evaluation of the one-loop determinant 
of the Chern-Simons gauge field in $AdS_3$.  Appendix D derives certain sums
which are used for evaluating the one-loop determinants in handlebody geometries.
Appendix E contains a description of the Mathematica files used to generate 
plots of the R\'{e}nyi entropies and the high temperature expansions.

\section{Bosonic twist correlators on the torus}

The  replica trick reduces the evaluation of the Renyi entropy for
a single interval in 
 $1+1$ CFT's   to  the computation of the two point function of twist fields. 
The CFT of interest is the 
 free complex boson CFT in $1+1$ dimensions, the complex boson is compactified
on square  torus of radius $\R$.  It obeys the boundary conditions
\begin{equation}\label{stbc}
 \phi(e^{2\pi i } z, e^{-2\pi i} \bar z) = \phi(z, \bar z ) + \R( m_1 + i m_2) , 
\end{equation}
The system is of finite size and is kept at finite temperature $\beta$. 
Let $\phi_i$ with $i=0,1, \cdots , n-1$ label the replica copies. To diagonalize the 
action of the cyclic permutation of the copies we consider the following 
linear combinations
\begin{equation} \label{fourier}
\tilde\phi_k = \sum_{j= 0}^{n-1} e^{2\pi i j \frac{k}{n} } \phi_j, 
\qquad k = 0, \cdots n-1. 
\end{equation}
$\tilde\phi_k$ is multiplied by the phase $e^{2\pi i \frac{k}{n}}$ as one moves around the
twist operator. 
The boundary conditions in (\ref{stbc}) also need to be imposed on 
$\tilde \phi_k$. Thus we obtain the following boundary conditions on $\tilde\phi_k$
 \begin{equation} \label{fullbc}
 \tilde\phi_k( e^{2\pi i } z, e^{-2\pi i  }\bar z) = e^{2\pi i \frac{k}{n} }\tilde\phi_k(z, \bar z )   + 
 \R\sum_{j=0}^{n-1} e^{2\pi i j \frac{k}{n}} ( m_{j, 1} + i m_{j, 2}) . 
 \end{equation}
 Thus $\tilde\phi_k$ is compactified  on the lattice $\Lambda_{\frac{k}{n}}$ which is defined 
 as 
 \begin{equation}
\Lambda_{\frac{k}{n}} \equiv
\left\{ q = \R\sum_{j=0}^{n-1} e^{\frac{2 \pi i k j }{n} } ( m_{j, 1} + i m_{j,2} ) ; \; m_{j, 1}, m_{j, 2} \in {\bf Z} 
\right\}. 
\end{equation}
 We now split $\tilde\phi_k$ into the classical and the quantum part as 
 $\tilde \phi_k = \tilde\phi_k^{qu} + \tilde\phi_k^{cl}$. The two point function of the 
 twist fields then splits into the a contribution from the quantum part and the sum over
 classical saddle points.  The classical part obeys the boundary condition
 \begin{equation}
 \tilde\phi_k^{cl}( e^{2\pi i } z, e^{-2\pi i }\bar z)  = e^{2\pi i \frac{k}{n}} \tilde\phi_k^{cl}( z, \bar z)  + v, 
 \end{equation}
 where $v$ is a vector in the lattice $\Lambda_{\frac{k}{n}}$. 
 The quantum part obeys the boundary condition
 \begin{equation}
 \tilde\phi_k^{qu}( e^{2\pi i } z, e^{-2\pi i }\bar z) = e^{2\pi i \frac{k}{n}} \tilde\phi_k^{qu}(z, \bar z).  
 \end{equation}
Let $\sigma_{k, n}$ be the twist field which responsible for the boundary 
condition  in (\ref{fullbc})
and $\bar\sigma_{k, n}$ the anti-twist field. 
Let the two point function of these operators on the torus be given by 
\begin{equation}
Z_{k, n} = \langle \sigma_{k, n}(z_1, \bar z_1 ) \bar \sigma_{k, n}( z_2, \bar z_2) \rangle. 
\end{equation}
Then the R\'{e}nyi entropy on the torus is given by 
\begin{equation}
\exp( (1-n) S_n) = \prod_{k=0}^{n-1} Z_{k, n} 
= \prod_{k=0}^{n-1} \langle \sigma_{k, n}(z_1, \bar z_1) \bar \sigma_{k, n}( z_2, \bar z_2) \rangle. 
\end{equation}

From this discussion, we see that the first task is to evaluate the two point function
of the twist operators on the torus.  This has been done in \cite{Atick:1987kd} and 
for the $n =2$ case by \cite{Saleur:1987tn}. 
To make our discussion self contained and simplified we review the procedure to 
evaluate this two point function. 
To un-clutter the notation, let the complex boson be 
denoted by  $X$. Then the steps to obtain the two point
function are as follows. 
\begin{enumerate}
\item We first write down the Greens functions
\begin{eqnarray}
g(z, w; z_i, \bar z_i) &\equiv & \frac{1}{Z_{k, n}} 
\langle - \partial_z X \partial_w \bar X \sigma_{k, n}(z_1, \bar z_1) \bar\sigma_{k, n} ( z_2, \bar z_2) 
\rangle, \\ \nonumber
h(\bar z, w; z_i, \bar z_i) &\equiv & 
\frac{1}{Z_{k, n}}  \langle - \partial_{\bar z} X \partial_w \bar X 
\sigma_{k, n}(z_1, \bar z_1) \bar\sigma_{k, n} ( z_2, \bar z_2) 
\rangle. 
\end{eqnarray}
where $z_i, i=1, 2$ are the two points at which the twist operators are inserted.  
These correlators are determined using the behaviour as $z, w$ approach the points at which 
the twist fields are inserted and monodromy conditions. 
\item The monodromy conditions are given as follows. 
First split the $X$ into a classical piece and a quantum fluctuation
\begin{equation}
 X = X_{cl} + X_{qu}. 
\end{equation}
When integrated over closed loops  over the two cycles of the torus 
the quantum piece does not change, that is 
\begin{equation} \label{mon1}
 \Delta_{\gamma_a} X_{qu} = \oint _{\gamma_a} dz \partial_z X_{qu} 
+ \oint_{\gamma_a} d \bar z \partial_{\bar z} X_{qu} = 0. 
\end{equation}
Here $\gamma_a, a=1, 2$ are the two cycles of the torus. 
The classical part obeys 
\begin{equation}\label{mon2}
 \Delta_{\gamma_a} X_{cl} = \oint _{\gamma_a} dz \partial_z X_{cl} 
+ \oint_{\gamma_a} d \bar z \partial_{\bar z} X_{cl} = v_a, 
\end{equation}
where $v_a$ is related to a translation on the compact  torus $\Lambda_{\frac{k}{n}}$, 
which will be specified 
subsequently. 
\item  The Green's functions $g(z, w; z_i, \bar z_i), h(\bar z, w; z_i, \bar z_i)$ 
are written down which obey the monodromy conditions and have the right singularity 
structure when $z, w$ approaches the twist insertions $z_i, \bar z_i$. 
 Then we obtain the expectation value
of the stress tensor by taking the following limit
\begin{equation}\label{stress}
 \langle T(z) \rangle = \lim_{\omega\rightarrow z } \left( g(z, w) - \frac{1}{( z - w)^2} \right). 
\end{equation}
\item
The OPE of the stress tensor with the primary field $\phi$ of weight $h_{\phi}$ is given by 
\begin{equation}
T(z) \phi(\omega, \bar\omega) \sim 
\frac{h_\phi}{( z-w)^2} + \frac{1}{(z-w)} \partial_w \phi( w, \bar w). 
\end{equation}
Using this OPE we can obtain the following set of differential equations for the 
two point function contribution from the quantum part $Z^{qu}_{k, n}$. 
\begin{equation}\label{wi}
 \partial_{z_i} \ln Z^{qu}_{k, n}  = \lim_{z\rightarrow z_i} 
\left( ( z-z_i)  \langle T(z) \rangle - \frac{h}{( z-z_i) } \right) , 
\end{equation}
where
\begin{equation}
h = \frac{1}{2}\frac{k}{n} \left( 1 - \frac{k}{n} \right), 
\end{equation}
is the weight of the twist field. 
This can be repeated and a similar set of differential equations can be obtained 
by considering the OPE of the anti-holomorphic stress tensor $\bar T(\bar z)$ with the
twist fields. 
\item
Solving these differential equations result in the quantum part of the 
two point function $Z^{qu}_{k, n}$. 
\item
The classical part of the two point function is obtained by considering the classical solutions
$\partial_z X_{cl}, \partial_{\bar z} X_{cl}$ 
which satisfy the monodromy conditions given in (\ref{mon1}).
We then evaluate  the classical 
action given by 
\begin{equation}
S_{cl} (X_{cl}, X_{cl} ) = \frac{1}{4\pi} \int d^2z ( 
\partial_z X \partial_{\bar z} \bar X + \partial_{\bar z} X \partial_{z} \bar X). 
\end{equation}
\item
The full two point function is then given by summing over all the classical solutions
\begin{equation}
 Z = Z^{qu}_{k, n} \sum_{\langle X_{cl} \rangle} e^{-S_{cl} }. 
\end{equation}
\end{enumerate}

\subsection{Quantum contribution}

In this subsection we construct the quantum contribution $Z^{qu}_{k, n}$ of the 
two point function of the twist operators. 
We begin with writing down the the Green's function $g$ and $h$  which satisfy the 
required conditions.
To do this we  write down the  cut differential for 
$\langle \partial_z X \rangle$, where the expectation 
value is taken in presence of the twist operators.  
The twist operator is located at $z_1$ while the anti-twist operator is 
located at $z_2$. 
 This is given by 
\begin{equation}
 \omega_1 (z)  = \vartheta_1( z- z_1)^{-(1-k/n) }\vartheta_1(z - z_2)^{-k/n} 
\vartheta_1( z -  (1-k/n) z_1 - (k/n)  z_2 ) . 
\end{equation}
Here $z_1$ is the location of the twist operator and $z_2$ is the location of 
the anti-twist operator. Note that these cut  differentials are 
doubly periodic, that is they are periodic under 
$z\rightarrow z+1$ and $z\rightarrow z+ \tau$  where $\tau$ is the modular parameter 
of the torus.  They also have the  required behaviour
\begin{eqnarray}
\lim_{z \rightarrow z_1} \langle \partial_z X \rangle 
&\sim &( z - z_1)^{ -( 1- \frac{k}{n} )} + \cdots, \\ \nonumber
\lim_{z\rightarrow z_2}  \langle \partial_z X \rangle 
&\sim &( z- z_2)^{ -k/n} + \cdots. 
\end{eqnarray}
We also have the cut differential for $\langle \partial_z \bar X \rangle$ which is 
given by 
\begin{equation}
 \omega_2 (z) = \vartheta_1( z-z_1)^{-k/n} \vartheta_1( z-z_2) ^{-(1-k/n) } 
\vartheta_1\left( z - \tfrac{k}{n} z_1 - \left( 1- \tfrac{k}{n}\right) z_2 \right) .  
\end{equation}
Again this differential is doubly periodic, and it has   the required behaviour
\begin{eqnarray}
\lim_{z\rightarrow z_1} \langle \partial_{\bar z}  X \rangle 
&\sim & ( z-z_1)^{-k/n}  + \cdots, \\ \nonumber
\lim_{z\rightarrow z_2} \langle \partial_{\bar z} X \rangle 
&\sim & ( z-z_2)^{-( 1-k/n)} + \cdots . 
\end{eqnarray}
The cut differentials  for 
$\langle \partial_{\bar z} X \rangle$ is given by  the complex conjugate
\begin{equation}
\bar\omega_2(\bar z) = \vartheta_1( \bar z-\bar z_1)^{-k/n} \vartheta_1( \bar z-\bar z_2) ^{-(1-k/n) } 
\vartheta_1\left( \bz - \tfrac{k}{n} \bz_1 - \left( 1- \tfrac{k}{n}\right) \bz_2 \right)  
\end{equation}
This is again doubly periodic and has the behaviour
\begin{eqnarray}
\lim_{\bar z \rightarrow \bar z_1} \langle \partial_{\bar z} X \rangle
&\sim & ( \bar z - \bar z_1)^{-k/n} + \cdots, \\ \nonumber
\lim_{\bar z \rightarrow \bar z_1} \langle \partial_{\bar z} X \rangle
&\sim &( \bar z-\bar z_2)^{-(1-k/n)} + \cdots .
\end{eqnarray}
Similarly the cut differential for $\langle \partial_{\bar z} \bar X \rangle$ 
is given by 
\begin{equation}
\bar \omega_1(\bar z) = 
\vartheta_1( \bar z- \bar z_1)^{-(1-k/n) }\vartheta_1(\bar z - \bar z_2)^{-k/n} 
\vartheta_1( \bar z -  ( 1-\tfrac{k}{n} )\bar z_1 - \tfrac{k}{n} \bar  z_2 ) . 
\end{equation}
This has the behaviour
\begin{eqnarray}
\lim_{\bar z \rightarrow \bar z_1} \langle \partial_{\bar z} X \rangle
&\sim & ( \bar z - \bar z_1)^{-(1-k/n)} + \cdots, \\ \nonumber
\lim_{\bar z \rightarrow \bar z_1} \langle \partial_{\bar z} X \rangle
&\sim & ( \bar z-\bar z_2)^{-k/n} + \cdots. 
\end{eqnarray}
Let us also define
\begin{eqnarray}
\gamma_{1-k/n} &=& \vartheta_1( z- z_1) ^{-(1-k/n) } \vartheta_1( z-z_2) ^{-k/n)},  \\ \nonumber
\gamma_{k/n}  &=& \vartheta_1( z-z_1)^{-k/n} \vartheta_1(z-z_2)^{-(1-k/n) },  \\ \nonumber
 Y_1 &=& -(1- \tfrac{k}{n})  ( z_1 - z_2) , \\ \nonumber
Y_2  &=&   \tfrac{k}{n}( z_1 - z_2) . 
\end{eqnarray}
to un-clutter our expressions. 

We can now write down the Green's functions. 
\begin{eqnarray}
g(z, w) &=& g_s( z, w) + A \omega_1(z) \omega_2(w) , \\ \nonumber
h(z, w) &=& B \bar \omega_2( \bar z ) \omega_2( w) .  
\end{eqnarray}
$A, B$  will be determined by the monodromy conditions given in (\ref{mon1}). 
$g_s$ contains the double pole in the respective Green's function and is given by 
\begin{equation}
 g_s(z, w ) =  \gamma_{1-k/n}(z) \gamma_{k/n}(w)
 \left( \frac{\vartheta_1'(0)}{\vartheta_1(z-w)} \right)^2 P(z, w) , 
\end{equation}
where 
\begin{eqnarray} \label{defp}
P(z, w) &=& \frac{k}{N} F_1(z, w)  \vartheta_1( w -z_1) \vartheta ( z- z_2) 
+ ( 1- \frac{k}{N}) F_2( z, w) \vartheta_1( w- z_2) \vartheta ( z- z_1),  \nonumber \\ 
F_1( z, w) &=& \frac{\vartheta_1( z -w + U_1)}{ \vartheta_1( U_1)}
\frac{\vartheta_1( z- w + Y_1 - U_1) }{\vartheta_1( Y_1 - U_1) }, \\ \nonumber
F_2( z, w) &=& \frac{\vartheta_1( z -w + U_2)}{ \vartheta_1( U_2)}
\frac{\vartheta_1( z- w + Y_2 - U_2) }{\vartheta_1( Y_2 - U_2) }. 
\end{eqnarray}
Note that $g_s( z, w) , P(z, w)$ is doubly periodic and $g_s$ has a double pole
at $z=w$ with coefficient $1$. 
$U_1, U_2$ are determined by the requirement that the single pole is 
absent in $g_s$ when $z\rightarrow w$. 
After some simple algebra it is easy see from the construction of $P(z,w)$
that this requirement is satisfied if 
$F_1$ and $F_2$ obey  the conditions
\begin{equation}
\partial_z F_1(z, w) |_{z=w} = 0, \qquad 
\partial_z F_2(z, w) |_{z=w} = 0. 
\end{equation}
This leads to the equations 
\begin{align} \label{zeros}
\frac{\vartheta'_1(U_1)}{\vartheta_1(U_1)}  + \frac{\vartheta'_1(Y_1 - U_1)}{\vartheta_1( Y_1- U_1)} 
&=0, \\ \nonumber
\frac{\vartheta'_1(U_2)}{\vartheta_1(U_2)}  + \frac{\vartheta'_1(Y_2 - U_2)}{\vartheta_1( Y_2- U_2)} 
&=0. 
\end{align}
Note that LHS of each equation is a doubly periodic 
meromorphic function of $U_1, U_2$ respectively.  
There is a simple pole at $U_1=0$ and $Y_1-U_1=0$ for the first LHS of the 
first equation. Similarly LHS of the second equation has a simple pole 
at $U_2 =0$ and $Y_2 -U_2 =0$. Therefore each of these functions has two zeros 
say at $U_1^{(0)}$ and $Y_1 - U_1^{(0)}$ and similarly for the second function. 
We will choose $U_1 = U_1^{(0)}$ and $U_2 = U_2^{(0)}$. Then the equations 
in (\ref{zeros}) are satisfied identically. 
We will not need the explicit solution $U_i^{(0)}$ or even the explicit form of 
$P(z, w)$ for evaluating the quantum part of the two point function. 
From the constraints on  $P(z, w)$  imposed by the fact that $g_s(z, w)$ has a double 
pole with coefficient $1$ and no single pole, it is easy to see that 
$P(z, w)$ obeys the following equation
\begin{align} \label{identa}
&\gamma_{1-k/n} (w) \gamma_{k/n}(w) \left(
\frac{\partial}{\partial z} \frac{\partial}{\partial w} P(z, w) |_{z=w }
+ \frac{\partial^2 }{\partial z^2} P ( z, w ) |_{z=w} \right) \nonumber \\ 
& \qquad = \left( 1 - \frac{k}{n} \right) \frac{ \vartheta_1''( w- z_1) }{ \vartheta_1( w - z_1) }
+ \frac{k}{n} \frac{\vartheta_1''( w - z_2) }{ \vartheta_1(w - z_2)} 
+ \frac{ \vartheta_1'( w- z_1)  \vartheta_1'( w - z_2) }{ \vartheta_1( w- z_1)  \vartheta_1( w - z_2) }.
\end{align}
We now determined $A, B$ from  the monodromy conditions given in (\ref{mon1}). 
We define the following integrals along the two cycles of the torus.
\begin{align} \label{cutint}
 W^1_1 = \int_{\gamma^1} dz \omega_1 (z) ,  \qquad
& W^2_1 = \int_{\gamma^1} d\bar z \bar\omega_2(\bar z) , \\ \nonumber
W^1_2 = \int_{\gamma^2} dz \omega_1( z) , \qquad
& W^2_2 = \int_{\gamma^2} d\bar z \bar \omega_2( \bar z) . 
\end{align}
Then the equations (\ref{mon1}) reduce to 
\begin{eqnarray}
  A \omega_2(w)   W^1_1 + B \omega_2(w) W^2_1 &=& - \int_{\gamma^1}  dz g_s( z, w) , \\ \nonumber
A \omega_2( \omega) W^1_2 + B \omega_2(w) W^2_2 &=& - \int_{\gamma^2}  dz g_s( z, w) .
\end{eqnarray}
The solution for $A, B$ are given by 
\begin{eqnarray}
 A \omega_2(w)  &=& - \frac{1}{|W|} \left( W^2_2  \int_{\gamma^1}  dz g_s( z, w) 
- W^2_1  \int_{\gamma^2}  dz g_s( z, w)  \right) , \\ \nonumber
B  \omega_2(w) &=& - \frac{1}{|W|} \left( W^1_1  \int_{\gamma^2}  dz g_s( z, w) 
- W^1_2  \int_{\gamma^1}  dz g_s( z, w)  \right) ,  \\ \nonumber
|W| &=& W^1_1 W^2_2 - W^1_2 W^2_1.
\end{eqnarray}
Note that these are just functions of $w$. 
Therefore the Greens' functions are 
\begin{eqnarray}
g(z, w)  &=& g_s(z, w) -\omega_1(z ) \frac{1}{|W|} \left( W^2_2  \int_{\gamma^1}  dy g_s( y, w) 
- W^2_1  \int_{\gamma^2}  dy g_s( y, w)  \right) , \nonumber \\ 
h(\bar z, w) &=& - \bar \omega_2(\bar z) \frac{1}{|W|} \left( W^1_1  \int_{\gamma^2}  dy g_s( y, w) 
- W^1_2  \int_{\gamma^1}  dy g_s( y, w)  \right). 
\end{eqnarray}

We  proceed to the second stage and obtain the stress tensor. Taking the limit
given in (\ref{stress}) and using the identity (\ref{identa})  we obtain
\begin{align}
 \langle T(z) \rangle =& \frac{1}{2} \left[ 
\frac{k}{n} \frac{\vartheta_1'( z - z_1) }{ \vartheta_1'( z-z_1) }
+ \left( 1- \frac{k}{n} \right)   \frac{\vartheta_1'( z - z_2) }{ \vartheta_1'( z-z_2) }
\right] 
\left[ \left( 1- \frac{k}{n} \right)\frac{\vartheta_1'( z - z_1) }{ \vartheta_1( z-z_1) }
+ \frac{k}{n} \frac{\vartheta_1'( z - z_2) }{ \vartheta_1'( z-z_2) } \right] \nn \\ \nonumber
&  - \frac{1}{2} \gamma_{1- \frac{k}{n} }(z) \gamma_k(z) 
\frac{\partial}{\partial z} \frac{\partial}{\partial w} P(z, w) |_{w = z}
- \frac{1}{3} \frac{\vartheta_1'''(0)}{\vartheta_1'(0)} \\  
&  - \omega_1(z ) \frac{1}{|W|} \left( W^2_2  \int_{\gamma^1}  dy \, g_s( y, z) 
- W^2_1  \int_{\gamma^2}  dy \, g_s( y, z)  \right).
\end{align}
We now  obtain differential equations for the quantum part of the 
two point function using the identity in (\ref{wi}). 
A similar set of differential equations can be obtained by considering 
the expectation value of the anti-holomorphic stress tensor in the 
presence of the twist fields.  We will examine the  differential equation
for $\partial_{z_1}\ln Z^{qu}_{k, n}$.  Taking the limit in  (\ref{wi}) we obtain 
\begin{align}
\partial_{z_1} \ln Z^{qu}_{k, n} =& 
\frac{1}{2} \left(  \Big(\frac{k}{n}\Big)^2 + \Big( 1 - \frac{k}{n}\Big)^2 \right) 
\frac{\theta_1' ( z_1 -z_2)}{\theta_1( z_1 - z_2) }  \\ \nonumber
&- \frac{1}{2 \theta_1'( 0 ) } \frac{1}{\vartheta_1( z_1 - z_2) } 
\frac{\partial}{\partial z}\frac{\partial}{\partial w} P(z, w) |_{z=w= z_1}  \\ \nonumber
&- \frac{1}{|W|} 
\left( W^2_2 \int_{\gamma^1} dz \omega_1(z) \Lambda (z) 
- W^2 _1 \int_{\gamma^2} dz \omega_1(z) \Lambda (z)  \right) ,
\end{align}
where
\begin{eqnarray}
 \Lambda(z) = \lim_{w \rightarrow z_1} \left( 
\frac{\omega_1( w) }{\gamma_{1- k/n} (w) } \right)
\frac{\gamma_{1-k/n}(z) }{\omega_1(z) }
\frac{ \vartheta_1'(0) P( z, z_1) }{ \vartheta_1^2( z- z_1) \vartheta_1(  z_1 - z_2)}  .
\end{eqnarray}
We will now rewrite $\Lambda$ in terms of the cut differential $\omega_1$. 
To do this we use  the form of $P(z, w)$ given in (\ref{defp}) and 
 the equations (\ref{zeros}) to Taylor expand $P(z, z_1)$ about $z_1$  and obtain
\begin{equation}
\lim_{z\rightarrow z_1} \Lambda(z) \sim \frac{1 - k/n}{z- z_1} 
\end{equation}
It is easy to see that this is also the leading singularity of the function 
$f(z) = \omega_1(z)^{-1} \partial_{z_1} \omega_1( z) $. 
The functions $\Lambda(z)$ and $f(z)$ are meromorphic doubly periodic functions
which agree at their poles. Therefore their difference must be a constant. 
Thus we have the equation
\begin{equation}\label{meoeq} 
 \omega_1(z) \Lambda(z)  = \partial_{z_1} \omega_1 + S \omega_1(z) 
\end{equation}
To obtain $S$, we examine the behaviour of the of 
$\Lambda(z) - ( \omega_1(z)) ^{-1} \partial_{z_1} \omega_1(z)$ as 
$z\rightarrow z_1$. This results in 
\begin{equation}
S = \frac{\partial_z^2 P(z, z_1)|_{z=z_1}}{ 2 \vartheta_1'(0) \vartheta_1( z_ 1 - z_2) }.
\end{equation}
We now substitute (\ref{meoeq}) and  use the identity (\ref{identa}) to write the differential equation
for the two point correlator as
\begin{equation}
\partial_{z_1} Z^{qu}_{k, n} = 
- \frac{1}{|W|} ( W^2_2 \partial_{z_1} W^1_1 - W^2_2 \partial_{z_1} W^1_2 ) 
+ \partial_{z_1} \ln \left[ \vartheta_1 ( z_1 - z_2) ^{-( 1- k/n) k/n} \right]. 
\end{equation}
Note that the first term can be rewritten as 
\begin{equation}
- \frac{1}{|W|} ( W^2_2 \partial_{z_1} W^1_1 - W^2_2 \partial_{z_1} W^1_2 ) 
= -\partial_{z_1} \ln |W|. 
\end{equation}
This is because the integrals $W^2_2, W^1_2$ depend on anti-holomorphic coordinates. 
Using this input,  and integrating the differential equation we obtain 
\begin{equation}
Z^{qu}_{k, n} = f(\tau, k/n, z_2 , \bar z_1, \bar z_2)  |W|^{-1} 
\left( \vartheta_1 ( z_1 - z_2) ^{-( 1- k/n) k/n} \right) .
\end{equation}
One can use the similar procedure to obtain a differential equation for $z_2$ and also 
for the anti-holomorphic coordinates $\bar z_1, \bar z_2$ by considering the 
expectation value of the the anti-holomorphic stress tensor $\langle \bar T(\bar z) \rangle$. 
This leads to 
\begin{equation}
Z^{qu}_{k, n} = f(\tau, k/n)  |W|^{-1}   \vartheta_1 ( z_1 - z_2) ^{-( 1- k/n) k/n} 
 \overline{ \vartheta_1 ( z_1 - z_2) }^{-( 1- k/n) k/n}.   
 \end{equation}
The normalization $f$ can be determined by  demanding that in the limit
 $z_1 \rightarrow z_2$, the two point function is normalized to one. 
 It is easy to see that in this limit  the integrals reduce to 
 \begin{align}
 W^1_1 \rightarrow \int_{0}^1dz =  1, \qquad & W^2_2 \rightarrow \int_0^\tau d\bar z = \tau, 
\\ \nonumber 
W^1_2 \rightarrow \int_0^{\bar \tau} dz =  \bar \tau, \qquad & W^2_1 \rightarrow \int_0^1 d\bar z =1. 
 \end{align}
Note that we are taking the cycle $\gamma_2$ to be such that $\bar z$ runs from $0$ to $\tau$, which 
implies $z$ runs from $0$ to $\bar\tau$. 
 Therefore $|W| \rightarrow 2 i  \tau_2$ where $\tau_2$ is the imaginary part of the modulus of the 
 torus. Using this argument we have
 \begin{equation}\label{quant} 
Z^{qu}_{k, n} = \frac{ 2 i \tau_2}{  |W|} 
\left(  \frac{ \vartheta_1'( 0)}{  \vartheta_1 ( z_1 - z_2)} \right) ^{( 1- k/n) k/n} 
 \left( \overline{ \frac{\vartheta_1'(0)}{ \vartheta_1 ( z_1 - z_2) } }\right)  ^{( 1- k/n) k/n}  .  
 \end{equation}
The quantum contribution of the Ashkin-Teller model \cite{Saleur:1987tn}
can be recovered  from the above expression by setting $k/n= 1/2$.

\subsection{Classical contribution}

\begin{figure}[!t]
\centering
\begin{tabular}{c}
\includegraphics[width=3.5in]{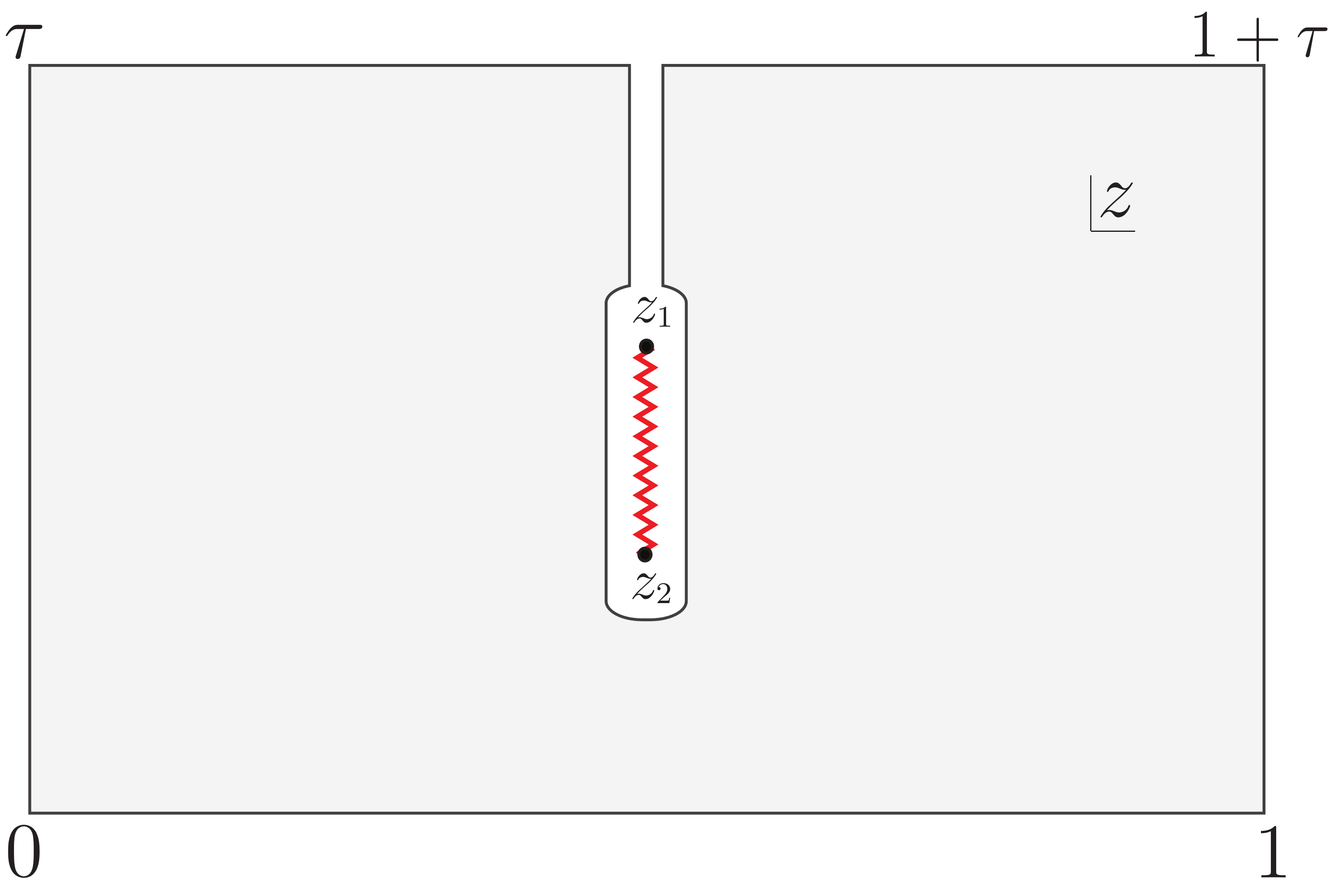} 
\end{tabular}
\caption{\small Simply connected domain of the area integral. }
\label{sal}
\end{figure}

In this sub-section we will evaluate the classical contribution due to the various saddle points
of the action. 
These saddle points exist when the complex boson is compactified on 
a torus. Consider the torus to be  a square torus of radius $\R + i\R$. 
Then the twisted sector boundary conditions 
for classical solutions is of the form
\begin{equation}
 X_\Cl ( e^{2\pi i } z, e^{-2\pi i} \bar z ) = e^{2\pi i k/n} X_{cl}( z, \bar z) +  v, 
\end{equation}
where $v$ is  a vector related to translations in the lattice $\Lambda_{\frac{k}{n}}$
which will be specified precisely subsequently.  
These boundary conditions give rise to the following 
monodromy conditions. 
\begin{align}\label{mono}
\oint _{\gamma_a} dz \, \pd _z X_\Cl + \oint_{\gamma_a} d\bz \, \pd_\bz X_\Cl &=  v_a \qquad (a=1,2) \nn \\
\oint _{\gamma_a} dz \, \pd _z \bX_\Cl + \oint_{\gamma_a} d\bz \, \pd_\bz \bX_\Cl &=  \bar v_a.
\end{align}
The subscript $a$ denotes the two cycles of the worldsheet torus. 
The classical action is given by 
\begin{align}\label{action}
S_\Cl [X, \bar{X}] = \frac{1}{4\pi} \int _D d^2 z \, ( \pd _z X_\Cl \pd _\bz \bar{X}_\Cl +  \pd _\bz X_\Cl \pd _z \bar{X}_\Cl ). 
\end{align}
We now have to evaluate the classical action given below for solutions satisfying the 
boundary conditions in (\ref{mono}).  The domain of integration $D$ will be specified
below. 
The classical solutions to the equations of motion can also be written in terms of the 
cut differentials. They are given by 
\begin{align}\label{sol-1}
\pd_z X_\Cl (z) &= a \omega_1(z),  \qquad \pd_\bz X_\Cl (\bz)= b \bar\omega_{2}(\bz),  \nn \\
\pd_z \bX_\Cl (z) &= \tilde a \omega_{2}(z),  \qquad 
\pd_\bz \bX_\Cl (\bz)= \tilde b \bar\omega_{ 1}(\bz). 
\end{align}
Note that these solutions satisfy the required holomorphic/anti-holomorphic conditions and the 
singularity properties near the twist operators. 
Substituting (\ref{sol-1}) in (\ref{mono}) we get the following system of equations. 
\begin{eqnarray}
\left( \begin{array}{cc}
        W^1_1 & W^2_1 \\
W^2_1 & W^2_2 
       \end{array}
\right)
 \begin{pmatrix}
a \\
b
\end{pmatrix} = \begin{pmatrix}
 v_1 \\
 v_2
\end{pmatrix} , \qquad 
\left( \begin{array}{cc}
        \bar W^2_1 & \bar W^1_1  \\
\bar W^2_2 & \bar W^1_2 
       \end{array}
\right)
 \begin{pmatrix}
\tilde a \\
\tilde b
\end{pmatrix} = \begin{pmatrix}
 \bar v_1 \\
 \bar v_2
\end{pmatrix}  . 
\end{eqnarray}
We have defined the  integrals of the cut differentials on the 
cycles of the worldsheet torus in (\ref{cutint}). 
The solutions for $a, \ b, \ \tilde{a}, \ \tilde{b}$ are as follows 
\begin{align}
a&=  \frac{W_2^2 v_1 -W_1^2 v_2}{|W|},  \qquad b=  \frac{-W_2^1 v_1 + W_1^1 v_2}{|W|},  \nn \\
\tilde{a} &= \frac{ \bar{W}_1^1 \bar{v}_2 - \bar W_2^1 \bar{v}_1}{|\bar{W}|},  \qquad 
\tilde{b} =  \frac{\bar{W}_2^2 \bar{v}_1 - \bar{W}_1^2 \bar{v}_2}{|\bar{W}|}. 
\end{align}
Substituting these in the solutions (\ref{sol-1}) and then evaluating the action (\ref{action})
we encounter the following area integrals. 
\begin{equation}
I_1 = \int d^2 z |\omega_1|^2 , \qquad I_2 = \int d^2 z |\omega_2|^2. 
\end{equation}
To perform these  integrals we use the  method developed in \cite{Saleur:1987tn} and 
\cite{Atick:1987kd}. 
Let us focus on the integral $I_1$. 
The area integral is carried out  over the shaded domain $d$ shown in figure \ref{sal}. 
In this domain, there exists an analytic function $f(z)$
\begin{equation}
f(z) = \int_{z_0}^z dt \, \omega_1(z), 
\end{equation}
such that  $df(z)  = \omega_1(z)$. 
Substituting this in the expression for $I_1$ we obtain
\begin{equation}
I_1 = \frac{1}{2i} \int dz \wedge \bar dz \, \frac{\partial}{\partial z} (  f(z) \bar \omega_1(\bar z )  ).  
\end{equation}
We can now use Green's theorem to write the area integral as an integral over the
contour whose boundary is the usual parallelogram representing the torus
along with the thin neck as shown in the Fig \ref{sal}.  
The integral along the cuts vanishes. This is because the combination $f(z) \bar \omega_1(z)$
does not jump across the cut.  The simplest way to see this is to take $z_0= z_1$ and then 
one sees that the monodromy of $\omega_1(z)$ cancels that of $\bar\omega_1(z)$. 
Thus the integral $I_1$ can be written as 
\begin{equation}
I_1 = \frac{1}{2i}(  \bar W^1_1 W^1_2 - W^1_1 \bar W^1_2) . 
\end{equation}
With a similar analysis we can show that 
\begin{equation}
I_2 = \frac{1}{2i} ( W^2_1 \bar W^2_2 - \bar W^2_1 W^2 _2) . 
\end{equation}
Now substituting these results in the classical action we obtain 
\begin{align}\label{act2}
S_{\Cl} =&  \frac{1}{4\pi \det(W)\det(\bar{W})} \Big( |v_1|^2 \im (W_2^2 W_2^1 \det(\bar{W}))	+ |v_2|^2 \im  (W_1^2 W_1^1 \det(\bar{W}))  \nn \\
&.\qquad\qquad\qquad\qquad\qquad
 - (v_1 \bar{v}_2 + \bar{v}_1 v_2) \im (W_2^1 W_1^2\det(\bar{W})) \Big). 
\end{align}
We will further simplify the classical action as well as the quantum contribution 
$Z^{qu}_{k, n}$ after we examine the integrals of the cut differentials. 

\subsection{Integrals of the cut differentials}
\begin{figure}[!t]
\centering
\begin{tabular}{c}
\includegraphics[width=3.5in]{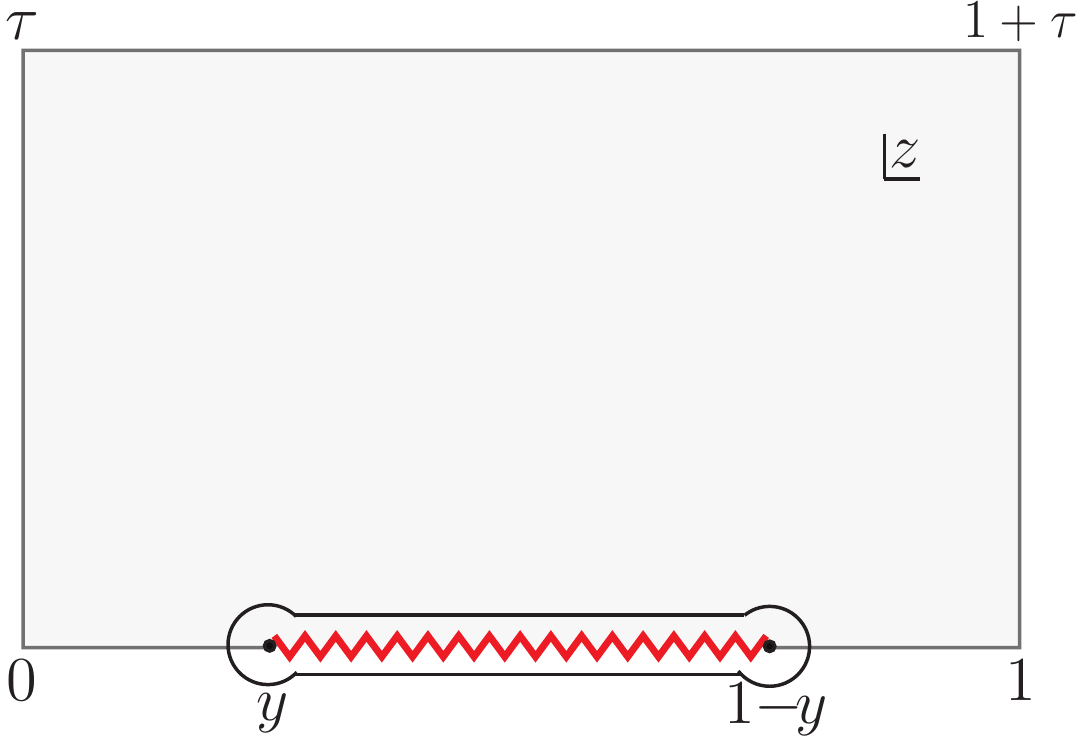} 
\end{tabular}
\caption{\small Fundamental domain of the torus lattice showing the location of the interval. }
\label{contour}
\end{figure}
There are four integrals of the cut differentials along  the two cycles of the 
worldsheet torus. 
To simplify the integrals further we choose  the points 
of insertions of the twist and the anti-twist operators to be at 
$z_1 = y$ and $z_2 = 1-y$ (where $0 \leq y \leq \tfrac{1}{2}$) as shown in Fig \ref{contour}. 
We will also choose the modular parameter of the worldsheet torus to be 
purely imaginary, that is 
\begin{equation}
\tau = i \beta.  
\end{equation}
From now onwards in this  paper we will work with these choices. 
Then the following properties of these integrals can be shown
\footnote{See appendix \ref{integral} for the proofs. }
\begin{align}\label{realitym}
(W_1^1)^* = W_1^2 ,\ \quad (W_1^1)^* = W_1^1 ,\ \quad (W_2^1)^* = -W_2^1 ,\ \quad (W_2^1)^* =  W_2^2. 
\end{align}
These identities imply that 
\begin{itemize}
\item \(W_1^1\) and \(W_1^2\) are purely real. 
\item \(W_2^1\) and \(W_2^2\) are purely imaginary.
\item \( \det (W)\) and \( \det(\bar{W})\) are purely imaginary. $\det(W) = 2i \,\im (W_1^1 W_2^2) $. 
\end{itemize}
Thus it is sufficient to examine the integrals $W_1^1$ and $W_2^2$. 
An important result which emerges during the proof of these relations is that 
the integral along the cut vanishes
\begin{equation}
\int_{y}^{1-y} dz \, \omega_1 (z) =0. 
\end{equation}
Thus the contribution of the $W^1_1$ integral arises only from the 
intervals $[0, y]$ and $[1-y, 1]$. Therefore we can write
\begin{equation}
W^1_1 = \int_0^y dz \, \omega_1(z) + \int_{1-y}^y dz \, \omega_1(z).  
\end{equation}

Let us define 
\begin{equation}
L = 1-2y,
\end{equation} 
the separation between the two insertions 
of the twist operator.
We will also keep track of  the dependence of the integrals 
on the interval $L$ and the modular  parameter of the worldsheet torus $\tau$. 
The integrals along the two cycles are  related by the following equation
\begin{equation}\label{cycle2}
W_2^2 ( L, \tau  ) = \tau W_1^1(  -L, \tau). 
\end{equation}
It can be seen from term by term expansion of  $W^1_1$ that 
\begin{equation}\label{parity}
W_1^1(L,\tau ) = W_1^1( -L, \tau). 
\end{equation}
Therefore we have
\begin{equation} \label{cycle3}
W_2^2(L,\tau ) = \tau W_1^1(  L, \tau). 
\end{equation}
Using modular transformation we can  obtain the high temperature 
expansions of the integral $W^2_2$ from the relation
\begin{equation}\label{mod}
W^2_2 ( L , \tau) =  \tau e^{\frac{i \pi  L^2  }{\tau} \frac{k}{n} ( 1- \frac{k}{n} )}  W^1_1 ( L/\tau, -1/\tau). 
\end{equation}
Once the high temperature expansion of $W^2_2$ is known, the high temperature 
expansion of $W^1_1$ can be obtained by the relation (\ref{cycle3}). 
All these relations are systematically proved in Appendix \ref{integral}. 
The  high-temperature expansion of  $W_2^2$ is derived in 
Appendix  \ref{integral} and is given by 
\begin{align}\label{hightemp3}
&W_2^2 (L,  \tau)_{k, n} \nn \\= & \; i e^{\frac{\pi}{\beta} \frac{k}{n} (1-\frac{k}{n})L^2}\beta \nn \\ & \times \Big{[} 1  -\frac{2}{n^2} {\left(-k^2+k n \cosh \left(\tfrac{2 \pi  L (n-k)}{n\beta}\right)+(k-n) \left(k \cosh \left(\tfrac{2 \pi  L}{\beta}\right)-n \cosh \left(\tfrac{2 \pi  k L}{n\beta }\right)\right)+k n-n^2\right)} q^2\nn \\
&\quad \ \ + \frac{1}{2 n^4}\left(2 n \left(3 (k-n) \left(k^2-k n+2 n^2\right) \cosh \left(\tfrac{2 \pi  k L}{n}\right) \right. \right. \nn \\ 
&\qquad\qquad \qquad \left. \left. +k \left(-3 \left(k^2-k n+2 n^2\right) \cosh \left(\tfrac{2 \pi  L (n-k)}{n}\right)-(k-2 n) (k-n) \cosh \left(\tfrac{2 \pi  L (k+n)}{n}\right) \right. \right.\right. \nn \\
&\qquad\qquad \qquad \qquad\qquad\left. \left. \left. +(k-n) (k+n) \cosh\left(\tfrac{2 \pi  L (k-2 n)}{n}\right)\right)\right) \right. \nn \\ 
&\qquad\qquad \left.-4 k (k-n) \left(k^2-k n+4 n^2\right) \cosh (\tfrac{2 \pi  L}{\beta})+3 \left(k^4-2 k^3 n+7 k^2 n^2-6 k n^3+4 n^4\right) \right. \nn \\ 
&\qquad\qquad +k (k-n) (k+n) (k-2 n) \cosh (\tfrac{4 \pi  L}{\beta})\Big{)} q^4 +\mathcal{O}(q^6) \Big{]}. 
\end{align}
The high temperature expansion 
of $W_1^1$ can then be obtained by the relation  in (\ref{cycle3}).

\subsection{R\'{e}nyi entropies of a compact boson}

We are now ready to put the results together to obtain the 
R\'{e}nyi entropies for a compact boson. 
The classical action in (\ref{act2}) simplifies on using the reality conditions to the following
\begin{align}\label{action1}
S_{cl} = \frac{1}{4\pi{\rm Im} ( \det( W) ) }  \left( |v_1 W_2^2|^2 	+ |v_2 W_1^1|^2 \right), 
\end{align}
We can now use the relation between $W^1_1$ and $W^2_2$ given in (\ref{cycle3})
to reduce the classical action to 
\begin{equation}\label{classical-action}
S_{cl} = \frac{1}{8\pi \tau_2} \left( |v_1|^2 |\tau|^2 + |v_2|^2 \right) . 
\end{equation}
Thus $S_{cl}$ is independent of the interval. This simplification is the result
of the  choice of the interval which is along one cycle of the 
torus as well as the fact that   the spacetime torus  is square 
and the worldsheet torus is rectangular. 
The lattice translations $v_1$ and $v_2$ are given as follows \cite{Calabrese:2009ez,Dixon:1986qv} 
\begin{equation}
v_j =  ( 1- e^{\frac{2\pi i k}{n} }) \xi_j, 
\end{equation}
and $\xi_j$ is  a general lattice vector in $\Lambda_{\frac{k}{n}}$. This lattice is defined as 
follows 
\begin{equation}
\Lambda_{\frac{k}{n}} \equiv
\left\{ q = \R\sum_{j=0}^{n-1} e^{\frac{2 \pi i k j }{n} } ( m_{j, 1} + i m_{j,2} ) ; \; m_{j, 1}, m_{j, 2} \in {\bf Z} 
\right\}. 
\end{equation}
Note that we are working with a square torus of radius $\R$ and this lattice results because of 
the Fourier transformation involved from the replica fields to the eigenbasis in which 
the shifts between replicas are represented by multiplication the phases $e^{\frac{2\pi i k}{n}}$
as given in (\ref{fourier}).   
Thus we can write
\begin{equation}
\xi_{p} = \R \sum_{j=0}^{n-1} e^{\frac{2 \pi i k j }{n} } ( m_{j, 1}^{(p)}  + i m_{j,2}^{(p)} ). 
\end{equation}
The superscript $(p)$ labels the two cycles of the worldsheet torus directions 
along which the monodromies are  evaluated.   The rest of the analysis proceeds
similar to the analysis done in \cite{Calabrese:2009ez}. 
We have
\begin{align}
|\xi_p|^2=\R^2 \left[ \sum_{q=1,2} [m_q^{(p)}]^t \cdot C_{k/n} \cdot m_q^{(p)} + \sum_{r,s=0}^{n-1} (m_{r,1}^{(p)}m_{s,2}^{(p)} - m_{s,1}^{(p)}m_{r,2}^{(p)} )  (S_{k/n})_{rs}\right]. 
\end{align}
Here, $m_q^{(p)} \in \mathbf{Z}$ and
\begin{align}
(C_{k/n})_{rs} \equiv \cos \left[ 2 \pi \frac{k}{n}(r-s) \right] \ , \quad (S_{k/n})_{rs} \equiv \left[ 2 \pi \frac{k}{n}(r-s) \right]. 
\end{align}
The classical action \eqref{classical-action} can be written in terms of $\xi_p$ as
\begin{equation}
S_{cl} = \frac{\sin ^2 (\tfrac{\pi k}{n})}{2\pi \tau_2} \left( |\xi_1|^2 |\tau|^2 + |\xi_2|^2 \right). 
\end{equation}
Thus the classical configurations are labelled by the integers $ m^{(p)}_{j} $ and we have to sum 
over these classical configurations. The contribution of the sum over classical configurations to the two point functions of the 
twist operator is therefore given by 
\begin{align}
Z_{cl} &= \prod_{k=0}^{n-1} Z_{k, n}^{cl},  \nonumber \\
&= \left[  \sum_{m \in \mathbf{Z}^n}\prod_{k=0}^{n-1} \exp \Big{\lbrace} - \frac{(1+|\tau|^2)\sin^2 (\tfrac{\pi k}{n})}{2\pi \tau_2} m^t \cdot C_{k/n} \cdot m  \Big{\rbrace} \right] ^2 ,  \nn \\
&= \left[\sum_{m \in \mathbf{Z}^n} \exp \Big{\lbrace}  i \pi [m^t \cdot \Omega  \cdot m ] \Big{\rbrace}   \right]^2 \label{pre-sum}
\end{align}
where we have defined 
\begin{align}
\Omega_{rs} = \frac{i(1+|\tau|^2 )\R^2}{2\pi \tau_2}\sum_{k=0}^{n-1} \sin^2 \left( \frac{\pi k}{n} \right) \cos \left( \frac{2\pi k }{n} (r-s) \right), 
\end{align}
where the indices $r$ and $s$ run over $1,2,\cdots,n$. The sum in \eqref{pre-sum} is however divergent since $\Omega$ has a zero eigenvalue. Following \cite{Calabrese:2009ez}, we can absorb this divergence in the normalization factor. Essentially this leads to fact that 
$r, s $ run from $1, 2, \cdots n-1$. Thus the partition function can be 
written in terms of 
 the Riemann-Siegel function as 
\begin{align}\label{class} 
Z_{cl} = \left[  
\Theta(0|\eta\Gamma)  \right]^2. 
\end{align}
Here
\begin{equation}
 \Theta(0|\eta\Gamma) =\sum_{m \in \mathbf{Z}^{n-1}} \exp ( i \pi m^t \cdot \Gamma \cdot m ), 
\qquad  \eta=\frac{i}{2}\frac{n(1+|\tau|^2)}{2\pi  \tau_2}\R^2 , 
\end{equation}
 and 
\begin{align}\label{cartan}
\Gamma_{rs} &= \frac{4}{n} \sum_{k=1}^{n-1}  \sin^2 \left( \frac{\pi k}{n} \right) \cos \left( \frac{2\pi k }{n} (r-s) \right)\nn  \\
&=2 \left(  \delta_{r,s} -\frac{1}{2}\delta_{r,s+1}-\frac{1}{2}\delta_{r,s-1}  \right). 
\end{align}
and $r, s$ run from $1, 2, \cdots n-1$. 
The above sparse matrix corresponds the Cartan matrix of $SU(n)$. The Riemann-Siegel function is then the partition function of the $A_{n-1}$  root lattice. This can be conveniently  expressed as the $\vartheta_3$-series  \cite{sloane1999sphere}.
The Riemann-Siegel function for the matrix $\Gamma$ given by \eqref{cartan} is thus given by 
\begin{align}
\Theta(0|\eta \Gamma) =  \frac{\sum_{k=0}^{n-2} 	\vartheta_3 (\tfrac{k}{n-1} | \eta)^{n-1}  }{ (n-1) \, \vartheta_3 (0|(n-1)\eta  )   } . 
\end{align}
The classical part \eqref{class}  upto a normalization reduces to 
\begin{align}
Z_{cl} = \left[   \frac{\sum_{k=0}^{n-2} 	\vartheta_3 (\tfrac{k}{n-1} | \eta)^{n-1}  }{ (n-1) \, \vartheta_3 (0|(n-1)\eta  )   }  \right]^2. 
\end{align}
It can be easily seen from the above that in the decompactification regime $\R \rightarrow \infty$, $Z_{cl}=1$.  The fact that we obtain the Riemann-Siegel theta function corresponding to the 
$A_{n-1}$ root lattice is a result of our special choice of the moduli. 
We have chosen a rectangular worldsheet torus as well as a square spacetime lattice. 
Furthermore the cut -- that is the location of the twist operators -- is along the 
spatial cycle of the torus. Due to these choices we obtained several simplifications
of the integrals of the cut differentials  which lead to the Riemann-Siegel theta 
function associated with $A_{n-1}$. For a generic choice of the torus moduli or 
the location of the twist operators we do not expect this simplification. 

\def\classical{{\left[   \frac{\sum_{k=0}^{n-1} 	\vartheta_3 (\tfrac{k}{n} | \eta)^n  }{ n \, \vartheta_3 (0|n\eta  )   }  \right]^2}}
\subsection*{The full partition function } 
We can now write the full partition function  taking contributions from the quantum \eqref{quant} and  classical  parts \eqref{class}
\begin{align}\label{full}
Z[n]=& \left( \prod_{k=0}^{n-1} 2 i  \tau_2 |W(k,n)|^{-1} \left(\frac{\vartheta_1 (z_2 - z_1|\tau)}{\vartheta'_1(0|\tau)}   \frac{\overline{\vartheta_1 (z_2 - z_1|\tau)}}{\overline{\vartheta'_1(0|\tau)}}\right)^{-\frac{k}{n}(1-\frac{k}{n})} \right) \nn \\ &\qquad \times  \left[   \frac{\sum_{k=0}^{n-2} 	\vartheta_3 (\tfrac{k}{n-1} | \eta)^{n-1}  }{ (n-1) \, \vartheta_3 (0|(n-1)\eta  )   }  \right]^2. 
\end{align}
The \Re entropy can be derived from the above by using 
\begin{align}
S_n = \frac{1}{1-n} \left( \log Z[n] - n \log Z[1] \right). 
\end{align}
Here, $Z[1]$ is the CFT partition function on the original spacetime (i.e.\,a torus in our case)  and $Z[n]$ is the partition function on the $\mathbb{Z}_n$ orbifolded cover. Using the above expression for $Z[n]$ (and $W(0,1)=2\tau$ for any $z_2 - z_1$, which is shown in \eqref{w01}), we can evaluate the \Re entropy to be  
\begin{align}\label{renyi-full}
S_{n} =&\frac{1}{1-n} \log \Biggr|    \left( \prod_{k=0}^{n-1}  \Big|\frac{W(k,n)}{2\tau}\Big|^{-1} \left(\frac{\vartheta_1 (z_2 - z_1)}{\vartheta'_1(0)} \frac{\overline{\vartheta_1 (z_2 - z_1)}}{\overline{\vartheta'_1(0)}}\right)  ^{-\frac{k}{n}(1-\frac{k}{n})}   \right) \Biggr|  \nn \\ 
&\qquad + \frac{1}{1-n} \log \Biggr|  \left[   \frac{\sum_{k=0}^{n-2} 	\vartheta_3 (\tfrac{k}{n-1} | \eta)^{n-1}  }{ (n-1) \, \vartheta_3 (0|(n-1)\eta  )   }  \right]^2   \Biggr| . 
\end{align}
We have thus obtained an exact analytical expression for the \Re entropies of the free boson on the torus.  It will be interesting to study this expression in detail and  especially explore the dependence
on  the radius $\R$. Compact bosons are related to several  $c=1$ statistical mechanical 
systems at criticality. Our result will be relevant to these systems. 

\section{R\'{e}nyi entropies for a non-compact boson}


For the rest of the paper we will  study the R\'{e}nyi entropies in the 
decompactification limit. In the 
$\R\rightarrow \infty$   only the quantum part contributes to the partition function.  Therefore the 
 full partition function (\ref{full}) reduces to 
\begin{align}
Z[n] = \left( \prod_{k=0}^{n-1}  \Big|\frac{W(k,n)}{2\tau}\Big|^{-1} \left(\frac{\vartheta_1 (z_2 - z_1)}{\vartheta'_1(0)} \frac{\overline{\vartheta_1 (z_2 - z_1)}}{\overline{\vartheta'_1(0)}}\right)  ^{-\frac{k}{n}(1-\frac{k}{n})}   \right) . 
\end{align}
The \Re entropy \eqref{renyi-full} is this regime is 
\begin{align}\label{renyi-1}
S_n =& -\frac{1}{1-n}     \sum_{k=0}^{n-1}   {\frac{k}{n}\left(1-\frac{k}{n}\right)}    \log   \Big{|} \frac{ \vartheta_1 (z_2 - z_1) \overline{\vartheta_1 (z_2 - z_1)} }{\vartheta'_1(0)^2}      \Big{|} -\frac{1}{1-n} \sum_{k=0}^{n-1} \log \Bigg{|} \frac{ W(k,n)}{2\tau}   \Bigg{|}.  
\end{align}

\subsection{Plots}

Although we have arrived at an expression for the \Re entropy, the answer hides two integrals (\ref{W-int})  involving $\vartheta$-functions.  As shown in appendix \ref{integral}, these integrals
can be performed in the high temperature or the low temperature expansion and analytical 
answers  for the R\'{e}nyi entropies can be obtained. 
However to get a picture of the behaviour of the R\'{e}nyi entropy as the function 
of the interval size and the temperature it is useful to 
 evaluate the integrals numerically and then plot the results. 

\begin{figure}[t]
\begin{tabular}{ll}
\includegraphics[scale=0.43]{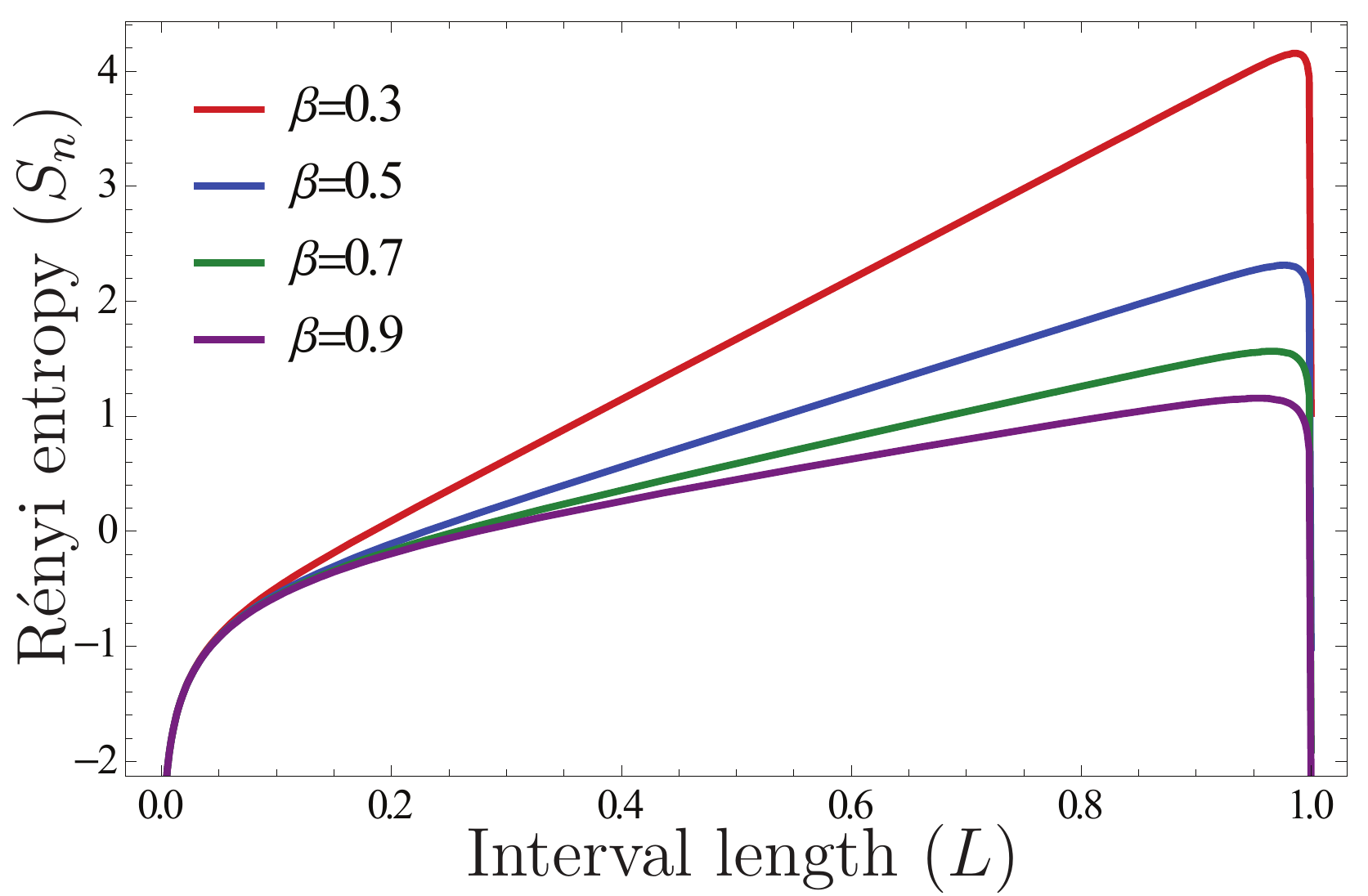}&\includegraphics[scale=0.43]{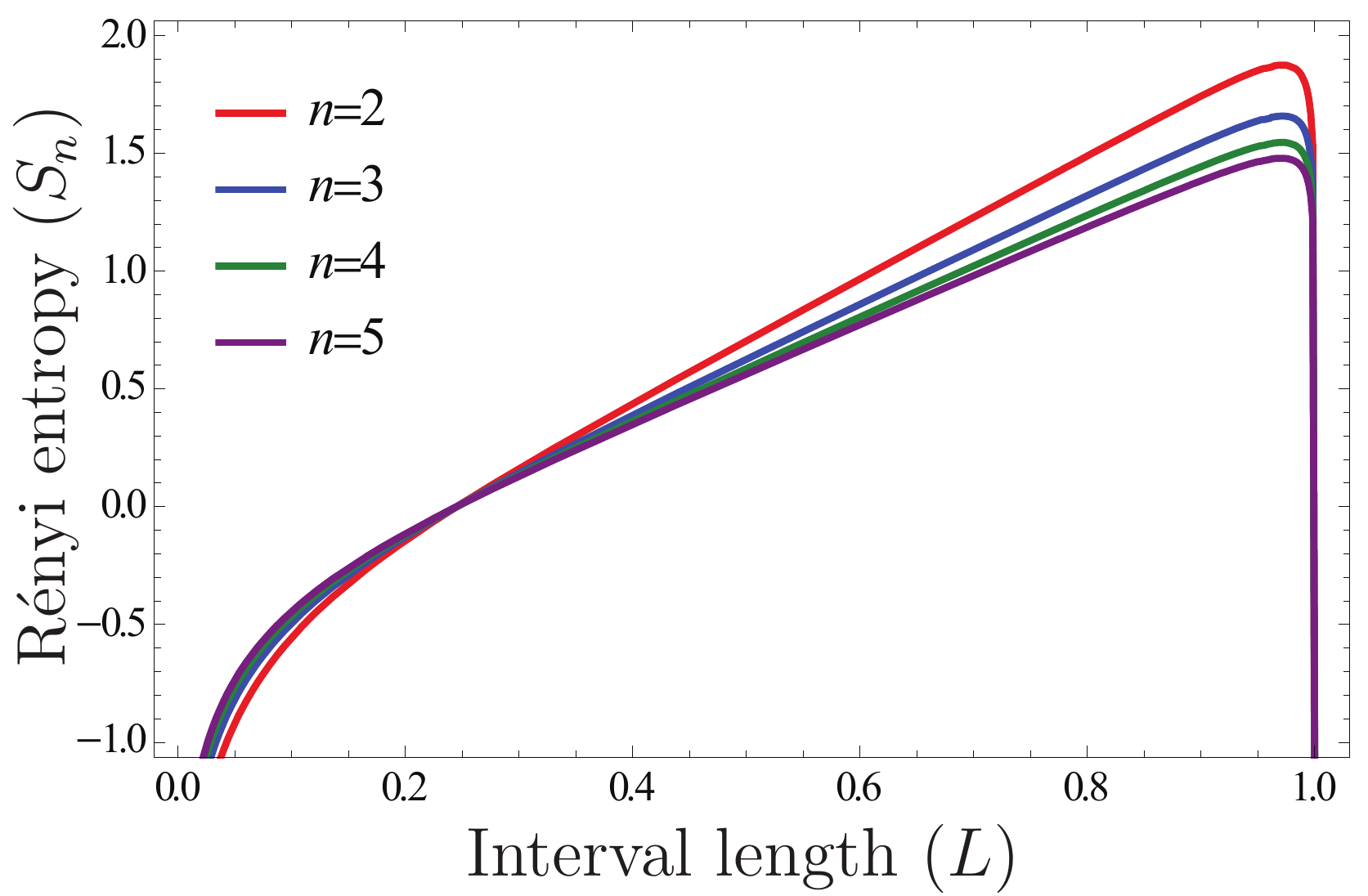}\\
\end{tabular}
\caption{\small (Left) Plots of the $n=2$ \Re entropies at $\beta = 0.3, 0.5, 0.7$ and $0.9$. (Right)  Plots of the $n=2, 3, 4$ and $5$ \Re entropies at $\beta = 0.6$. }
\label{renyi-plot}
\end{figure}

In Fig \ref{renyi-plot} we have plotted the $n=2$ \Re entropies at four different temperatures. It can be seen that as the temperature increases the curve shifts higher. Also, the \Re entropies for  $n=2, 3, 4$ and 5 are plotted for $\beta =0.6$. 
\begin{figure}[t]
\centering
\begin{tabular}{c}
\includegraphics[scale=0.44]{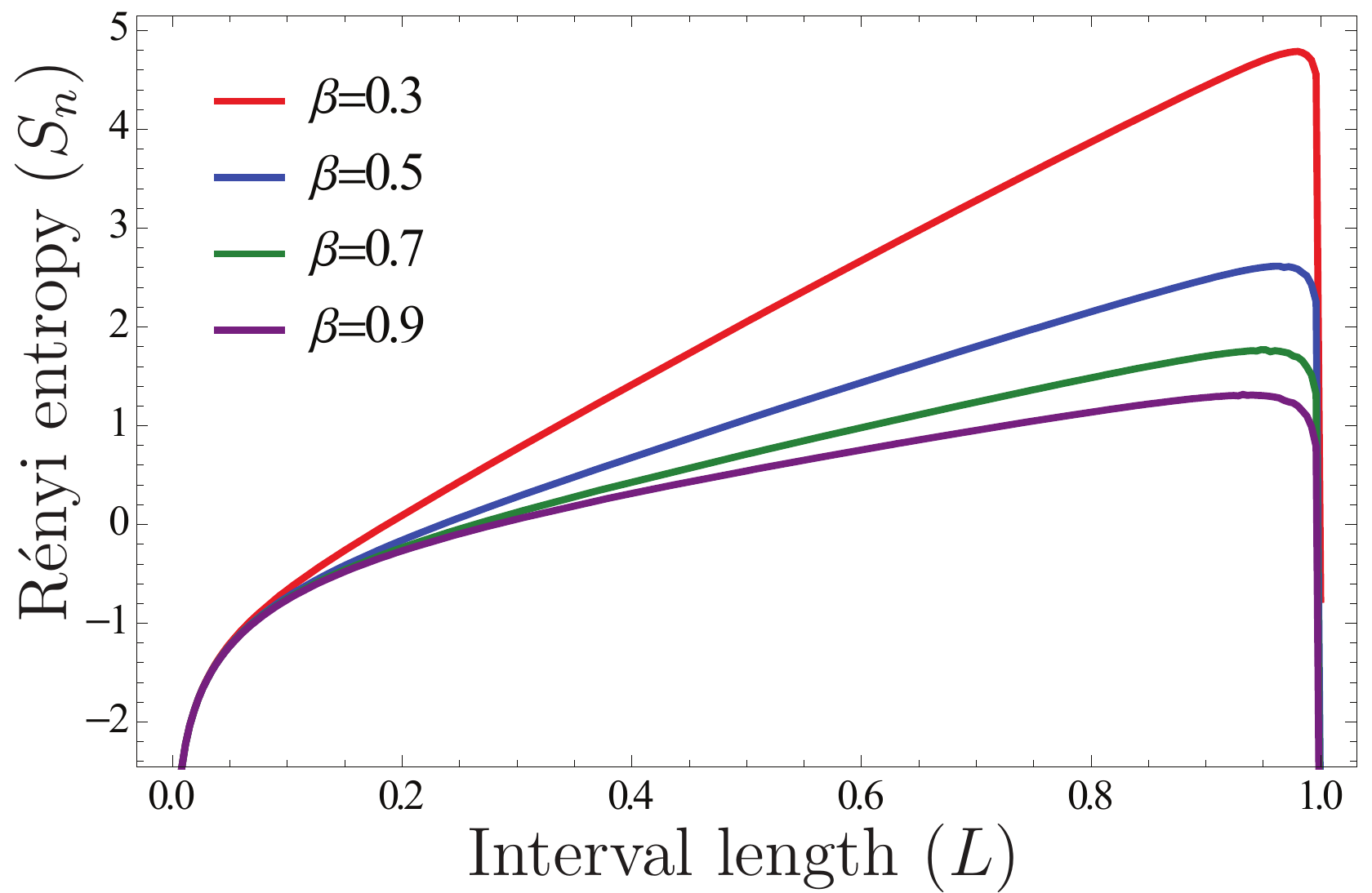} 
\end{tabular}
\caption{\small Plot of \Re entropy extrapolated to $n\rightarrow 1$ at $\beta = 0.3, 0.5, 0.7$ and $0.9$. \ \ \ \ \ \ \ \ \ \ }
\label{extra-plot}
\end{figure}
These plots are similar to those obtained for the case of fermions on the torus \cite{Azeyanagi:2007bj,Herzog:2013py}. 
We have also  numerically extrapolated the answers for the \Re entropies ($n=2,3,\cdots,10$) at four different temperatures to $n\rightarrow 1 $ and obtained the entanglement entropy. The plot is shown the Fig \ref{extra-plot}.
Note that for all these plots we have set the cut-off to unity. 

The details of the Mathematica files used to generate these plots is given in Appendix \ref{mathematica}. 

As it can be seen the plots, the \Re or entanglement entropy increases with rise in temperature. This can be reasoned as follows : increasing the temperature makes the system  accessible to more excited states and thereby increases the Hilbert space we are tracing over.  The peak at around 
$L=1$ is the result of the fact that at this point the entanglement entropy 
reduces to the thermal entropy which we now explore in the next subsection.

\subsection{Thermal entropy from entanglement entropy}

It is known that the entanglement entropy reduces to the thermal entropy when the size of the interval equals the length of the full spatial circle. Since there are no partial traces to perform,  it is then given by
the usual density matrix   $\rho = \frac{e^{-\beta H}}{Z}$. 
The entanglement entropy is then
\begin{align}
S_E & = \lim_{n\rightarrow 1}S^{(A)}_n = - \frac{\pd}{\pd n} \log (\text{Tr} (\rho ^n)) \Big|_{n=1} \nn\\
 &= -\frac{\pd}{\pd n} \left( \log (\text{Tr}(e^{-n \beta H }))-n \log Z \right)  \Big|_{n=1}  \nn \\  
&=\beta \langle H \rangle + \log Z = \beta ( E- F)   = S_{\text{thermal}}, 
\end{align}
 here `Tr' indicates not the partial trace but the trace over the full Hilbert space of states. 
In this section we shall calculate the entanglement entropy in this limit and verify that it does match with the thermal entropy.  A similar check for the case  of fermions on the torus  was done in \cite{Azeyanagi:2007bj}.

\subsection*{Entanglement entropy in the limit $y\rightarrow 0$}
Let's define
\begin{align}\label{ee-1}
S_1 &=  -\frac{1}{1-n}     \sum_{k=0}^{n-1}   {\frac{k}{n}\left(1-\frac{k}{n}\right)}    \log   \Big{|} \frac{  \vartheta_1 (z_2 - z_1) \overline{\vartheta_1 (z_2 - z_1)}   }{\vartheta'_1 (0)^2}    \Big{|}, \nn
\\
& = \frac{n+1}{6n} \log \Big{|} \frac{ \vartheta_1 (z_2 -z_1)}{\vartheta'_1 (0)^2}    \Big{|}^2 ,   \\
S_2 &= -\frac{1}{1-n} \sum_{k=0}^{n-1} \log \Bigg{|}  \frac{W(k,n)}{2\tau}   \Bigg{|}. 
\end{align}
It follows  from these definitions that $S_n = S_1 + S_2 $. 
We shall take the limit $y\rightarrow 0$ in which the length of the  interval ($L=1-2y$) equals the size of the full spatial circle or the system size (see Fig \ref{contour}). 
We can then extract the finite part of the entanglement entropy as follows
\begin{align}\label{finite}
S_{\text{finite}} (L\rightarrow 1)= S(1-\epsilon) - S(\epsilon). 
\end{align}
In this section, the interval length is denoted as $L$ and the full spatial circle ($R$) is taken to be unity.

From (\ref{ee-1}) it can be seen that $S_1$ captures the universal piece of the entanglement entropy for $n\rightarrow 1$. The high-temperature expansion for $S_1$ can be obtained by performing a S-modular transformation $\tau \rightarrow -1/\tau$ in equation (\ref{ee-1}) 
\begin{align}
S_1 (L)\n1 &=  \frac{2}{3} \log \Big{|} \frac{ \vartheta_1 (L|\tau)}{2\pi \eta(\tau)^3}    \Big{|},  \nn \\ &=  \frac{2}{3} \log \Big{|}(-i\tau) e^{-\pi L^2/\beta} \frac{ \vartheta_1 (\tfrac{L}{\tau}|-\tfrac{1}{\tau})}{2\pi \eta(-\frac{1}{\tau})^3}    \Big{|},  \nn \\
&= \frac{2}{3} \log \Big|\frac{ \beta}{2\pi} \sinh\left( \frac{\pi L}{\beta} \right) e^{-\pi L^2/\beta} \Big| \nn \\ &\quad + \frac{2}{3} \log \Big{|}  \prod_{m=1}^\infty \frac{(1-e^{-2\pi m/\beta})(1-e^{-2\pi L m/\beta}e^{-2\pi m/\beta})(1-e^{2\pi L m/\beta}e^{-2\pi m/\beta})}{(1-e^{-2\pi m/\beta})^3} \Big|. 
\end{align} 
In both the limits $L\rightarrow 1 - \epsilon$ and $L \rightarrow \epsilon$ this becomes 
\begin{align}
S_1 (1-\epsilon \text{ or } \epsilon)\n1 &= \frac{2}{3} \log \epsilon . 
\end{align}
So, there is no   contribution from $(S_1(1-\epsilon) - S_1(\epsilon))|_{n=1}$ to the finite part of the entanglement entropy (\ref{finite}) 

For $S_2$ we have 
\begin{align}\label{s2}
S_2 (L)&= -\frac{1}{1-n} \sum_{k=0}^{n-1} \log \Big{|}  \frac{W(k,n)}{2\tau}   \Big{|},   
\end{align}
with $\det (W) = 2i\, \im (W_1^1 W_2^2)$. 
The integrals of $W_1^1$ and $W_2^2$ in the limit $y\rightarrow 0$ are evaluated in Appendix \ref{integral}. We get  the following from \eqref{w1-y0} and \eqref{w2-y0}.
\begin{align}
W_1^1 =   \frac{\vartheta_1 ( - \tfrac{k}{n})}{\vartheta'_1 (0) } \ , \qquad W_2^2 =   \tau  \frac{\vartheta_1 ( - \tfrac{k}{n})}{\vartheta'_1 (0) }. 
\end{align}
\begin{align} 
\det (W(k,n)) &=2\tau \Big| \frac{\vartheta_1 (  \tfrac{k}{n})}{\vartheta'_1 (0) }  \Big| ^2 . 
\end{align}
The expression for $S_2(1)$ then becomes
\begin{align}
S_2 (1)= -\frac{1}{1-n} \sum_{k=0}^{n-1} \log  \Big| \frac{\vartheta_1 ( \tfrac{k}{n})}{\vartheta'_1 (0) }  \Big| ^2. 
\end{align}
Since we are interested in the entanglement entropy, we take the $n\rightarrow 1$ limit of the above expression. 
\begin{align}
S_2 (1) \n1 &= \frac{-1}{1-n} \log \Big|   \prod_{k=0}^{n-1} \left( \frac{\vartheta_1(k/n)}{\vartheta_1'(0)} \right) \Big|^2 \n1 \nn,   \\
&= \frac{\pd}{\pd n} \sum_{k=0}^{n-1}  \log \Big|  \frac{\vartheta_1(k/n)}{\vartheta_1'(0)} \Big|^2 \n1. 
\end{align}
On performing a S-modular transformation, since we are interested in the high temperature expansion we get
\begin{align}
S_2(1)\n1&= 2 \frac{\pd}{\pd n} \sum_{k=0}^{n-1}  \log \Big|  \frac{\vartheta_1(\tfrac{k}{n\tau}|-\tfrac{1}{\tau})}{\vartheta_1'(0|-\tfrac{1}{\tau})} \Big| \n1+ \frac{\pd}{\pd n} \sum_{k=0}^{n-1} \left( - \frac{2\pi i k^2}{n^2 \tau} \right)    \n1           \nn,  \\
&= - \frac{\pi}{3\beta} +  2 \frac{\pd}{\pd n} \sum_{k=0}^{n-1}  \log \Big|  \frac{\vartheta_1(\tfrac{k}{n\tau}|-\tfrac{1}{\tau})}{\vartheta_1'(0|-\tfrac{1}{\tau})} \Big| \n1. 
\end{align}
Let us now consider the second term above
\begin{align}
s_2 &= 2 \frac{\pd}{\pd n} \sum_{k=0}^{n-1}  \log \Big|  \frac{\vartheta_1(\tfrac{k}{n\tau}|-\tfrac{1}{\tau})}{\vartheta_1'(0|-\tfrac{1}{\tau})}  \Big|  \n1, 
 \nn \\
&= 2 \pd_n \sum_{k=0}^{n-1} \log \Big| \sinh \left(\frac{\pi k}{n \beta}\right) \prod_{m=1}^{\infty} \frac{(1-e^{\frac{2\pi k }{n\beta}}e^{-\frac{2\pi m }{\beta}})(1-e^{-\frac{2\pi k }{n\beta}}e^{-\frac{2\pi m}{\beta}})}{(1- e^{-\frac{2\pi m }{\beta}})^2} \Big| \n1,  \nn \\
&= 2 \pd_n \sum_{k=0}^{n-1} \log \Big| \sinh \left(\frac{\pi k}{n \beta}\right) \Big| + 2 \pd_n \sum_{k=0}^{n-1} \log \Big|  \prod_{m=1}^{\infty} \frac{(1-e^{\frac{2\pi k }{n\beta}}e^{-\frac{2\pi m }{\beta}})(1-e^{-\frac{2\pi k }{n\beta}}e^{-\frac{2\pi m}{\beta}})}{(1- e^{-\frac{2\pi m }{\beta}})^2} \Big| \n1. 
\end{align}
The second term in the last line of the above equation can be written as
\begin{align}
-8	\pd _n \sum_{k=0}^{n-1} \sum_{m=1}^{\infty} \sum_{l=1}^{\infty} \frac{1}{l} e^{-2\pi m l/\beta} \sinh ^2\left( \frac{\pi k l}{n \beta}  \right) \n1. 
\end{align}
Performing the sums over $m$ and $k$ we get
\begin{align}
&-8\pd _n \sum_{l=1}^\infty \frac{1}{l} \frac{e^{-2\pi l /\beta}}{1-e^{-2\pi l /\beta }} \frac{1}{4} \left(  \frac{\sinh(\tfrac{\pi l}{\beta n}(2n-1))}{\sinh (\tfrac{\pi l}{\beta n })} - 2n +1\right) \n1,  \nn \\
=& 4 \sum_{l=1}^\infty \frac{1}{l} \frac{e^{-2\pi l /\beta}}{1-e^{-2\pi l /\beta}} \left( 1- \frac{\pi l}{\beta} \cot (\tfrac{\pi l}{\beta})   \right). 
\end{align}
Therefore, the final expression for $S_2$ in the $n\rightarrow 1 $ limit is given by 
\begin{align}
S_2(1)\n1 &= - \frac{\pi}{3 \beta} + 2 \pd_n \sum_{k=0}^{n-1} \log \Big| \sinh (\frac{\pi k}{n \beta}) \Big| \n1 +4 \sum_{l=1}^\infty \frac{1}{l} \frac{e^{-2\pi l /\beta}}{1- e^{-2\pi l /\beta}} \left( 1- \frac{\pi l}{\beta} \cot (\tfrac{\pi l}{\beta})    \right),  \nn \\
&=  - \frac{\pi}{3 \beta} +  \left( \frac{\pi}{\beta} - \frac{4\pi}{\beta} \sum_{l=1}^\infty \frac{1}{e^{2\pi l /(n\beta)}-1}   \right)       +4 \sum_{l=1}^\infty \frac{1}{l} \frac{e^{-2\pi l /\beta}}{1- e^{-2\pi l /\beta}} \left( 1- \frac{\pi l}{\beta} \cot (\tfrac{\pi l}{\beta})    \right),  \nn \\
&=  \frac{2\pi}{3\beta} - \frac{4\pi}{\beta} \sum_{l=1}^\infty \frac{1}{e^{2\pi l /(n\beta)}-1}       +4 \sum_{l=1}^\infty \frac{1}{l} \frac{e^{-2\pi l /\beta}}{1- e^{-2\pi l /\beta}} \left( 1- \frac{\pi l}{\beta} \cot (\tfrac{\pi l}{\beta})    \right). 
\end{align}
It can be checked that for $L=\epsilon$ we have  $W(k,n)=\tau=W(0,1)$. 
Therefore, $S_2(\epsilon)$ is zero from (\ref{s2}). 

Thus, the finite contribution to the entanglement entropy (\ref{finite}) at $L\rightarrow 1$ is 
\begin{align}\label{ee-thermal-final}
S_{\text{finite}} (L\rightarrow 1) 
&=  \frac{2\pi}{3\beta} - \frac{4\pi}{\beta} \sum_{l=1}^\infty \frac{1}{e^{2\pi l /(n\beta)}-1}       +4 \sum_{l=1}^\infty \frac{1}{l} \frac{1}{ e^{2\pi l /\beta}-1} \left( 1- \frac{\pi l}{\beta} \cot \left(\frac{\pi l}{\beta}\right)    \right)  .
\end{align}

\def\osc{\text{osc}}
\def\kin{\text{kin}}
 \subsection*{Thermal entropy of free bosons on a torus}
We shall now calculate the thermal entropy for free bosons on a circle at finite temperature and verify whether it matches with (\ref{ee-thermal-final}).   For a free complex boson on a circle 
the contribution to the partition function from the oscillator modes is given by \cite{Ginsparg:1988ui}
\begin{align}\label{thermal-Z}
Z_\osc=\frac{1}{|\eta(\tau)|^4} . 
\end{align}
%
On performing a modular transformation we get 
\begin{align}
Z_\osc= \beta^{-2} e^{\frac{\pi}{3\beta}} \prod_{n=1}^\infty \Big| 1- e^{\frac{-2\pi n}{\beta} } \Big| ^{-4}. 
\end{align}
In addition to the oscillator modes we also need to take into consideration the contribution from the zero modes or the momenta. The contribution of these modes to the partition 
function is given by 
\begin{align}
Z_\kin = \left[ \int_{-\infty}^\infty dp_1 dp_2 e^{-\beta(p_1^2 +p_2^2)} \right] ^2 = \left( \frac{2\pi}{\beta} \right)^2. 
\end{align}
The full partition function is then given as a product 
\begin{align}
Z=Z_\osc Z_\kin. 
\end{align}
The thermal entropy can therefore be calculated as follows
\begin{align}
S  &= \beta^2 \pd_\beta (-\beta^{-1} \log Z),  \nn \\
&=  \frac{2\pi}{3\beta} -4 \sum_{n=1}^\infty \log\Big| 1- e^{\frac{-2\pi n}{\beta} } \Big| - \frac{8\pi}{\beta} \sum_{n=1}^\infty \frac{n }{e^{2\pi n  /\beta}-1}. 
\end{align}
The above expression matches with (\ref{ee-thermal-final}). 
This  can be seen from the following identities 
\begin{align}
\sum_{l=0}^\infty \frac{1}{l} \, \frac{1}{ 1- e^{-2\pi l \beta}} &= - \sum_{n=0}^\infty \log \Big| 1 - e^{-2\pi n /\beta}     \Big|,  \nn \\
 \sum_{l=1}^\infty \frac{1}{ e^{2\pi l /\beta}-1 } \cot \left(\frac{\pi l}{\beta}\right)  &= 2 \sum_{m=1}^\infty \frac{m}{e^{2\pi m /\beta} -1} -  \sum_{m=1}^\infty \frac{1}{e^{2\pi m /\beta} -1} \nn. 
\end{align}
This verifies the claim  that the entanglement entropy reduces to the thermal entropy in the limit of the interval size equalling the size of the full spatial circle.

\def\bz{{\bar{z}}}
\section{Finite size corrections to R\'{e}nyi entropies}

In this section we set up a high temperature expansion of the R\'{e}nyi entropies in order 
to extract finite size corrections to the universal part. 
We first do this for bosons and then for then for the  fermions.  
Let's define the following\footnote{These are same definitions as in \eqref{ee-1} but with the extra $(n)$ superscript.}
\begin{align}
S_n &= S_1^{(n)} + S_2^{(n)},  \nn \\
S_1^{(n)} &=  -\frac{1}{1-n}     \sum_{k=0}^{n-1}   {\frac{k}{n}\left(1-\frac{k}{n}\right)}    \log   \Big{|} \frac{  \vartheta_1 (z_2 - z_1) \overline{\vartheta_1 (z_2 - z_1)}   }{\vartheta'_1 (0)^2}    \Big{|} = \frac{n+1}{6n} \log \Big{|} \frac{ \vartheta_1 (z_2 -z_1)}{\vartheta'_1 (0) }    \Big{|}^2,  \label{s1}   \\
S_2^{(n)} &= -\frac{1}{1-n} \sum_{k=0}^{n-1} \log \Big{|}  \frac{W(k,n)}{2\tau}   \Big{|}.  
\end{align}
In order to evaluate the entanglement entropy we should take the $n\rightarrow 1$ limit of the above. We get
\begin{align}
S_1^{(1)} &= \frac{2}{3} \log \Big{|} \frac{ \vartheta_1 (z_2 -z_1)}{\vartheta'_1 (0)}    \Big{|},    \\
S_2^{(1)} &= \pd_n\left(  \sum_{k=0}^{n-1} \log \Big{|}  \frac{W(k,n)}{2\tau}   \Big|  \right) \Bigg|_{n=1}. 
\end{align}

\subsection*{High temperature expansions of \Re and entanglement entropies}
We are now interested in the high-temperature expansion. We shall retain just the leading order terms which will facilitate the comparison with the holographic computation. 
The high-temperature expansion of $W_2^2(k,n)$ is obtained in  \eqref{s21}. 
The interval of length $L(=1-2y)$ is now defined by $z_1=y=\frac{1-L}{2}$ and $z_2=1-y=\frac{1+L}{2}$.
The expressions for the \Re and entanglement entropy  are
\begin{eqnarray}
S^{(n)}_1  &=& \frac{n+1}{3n} \log \Big{|}   \frac{\beta}{\pi} e^{-\pi L^2 / \beta} \sinh \left( \frac{\pi L}{\beta} \right)     
\prod_{m=1}^\infty \frac{(1- e^{2\pi L /\beta}q^m)(1- e^{-2\pi L /\beta}q^m) }{(1-q^m)^2}
  \Big{|},      \\
  &=&  \frac{n+1}{3n} \log \Big{|}   \frac{\beta}{\pi} e^{-\pi L^2 / \beta} \sinh \left( \frac{\pi L}{\beta} \right) \Big|
   - \frac{n+1}{3n}  \sum_{m=1}^\infty \sum_{l=1}^\infty \frac{q^{ml}}{l}4\sinh^2 \left( \frac{\pi L l}{\beta}  \right),  \nn \\ \nn
&=& \frac{n+1}{3n} \log \Big{|}   \frac{\beta}{\pi} e^{-\pi L^2 / \beta} \sinh \left( \frac{\pi L}{\beta} \right) \Big|
   -\frac{4(n+1)}{3n} e^{-2\pi T} \sinh^2 \left({\pi L}{T} \right) + \mathcal{O}(e^{-4\pi T}) ,  \\
 \label{s11}
S_1^{(1)}    &=&   \frac{2}{3} \log \Big{|}   \frac{\beta}{\pi} e^{-\pi L^2 / \beta} \sinh \left( \frac{\pi L}{\beta} \right) \Big|
   -\frac{8}{3} e^{-2\pi T} \sinh^2 \left( {\pi L}{T} \right) 
+ \mathcal{O}(e^{-4\pi T }). 
\end{eqnarray}
Now using the fact that $\det(W)= 2i\text{Im}(W_1^1 W_2^2)$ and $W_2^2 = \tau W_1^1$ as 
shown in appendix \ref{integral}) we obtain
\begin{align}\label{s21}
S_2^{(n)} &=  - \frac{1}{1-n} \left(  \sum_{k=0}^\infty \log |W_1^1(k,n)| +    \sum_{k=0}^\infty \log \Big|\frac{W_2^2(k,n)}{\beta}\Big| \right),  \nn \\
 &= - \frac{1}{1-n}  \left(2  \sum_{k=0}^\infty \log \Big|\frac{W_2^2(k,n)}{\beta}\Big| \right) , \\
 S_2^{(1)} &= \pd_{n}\left(2  \sum_{k=0}^\infty \log \Big|\frac{W_2^2(k,n)}{\beta}\Big| \right).  \label{s2n}
\end{align}
The expression for  $W_2^2$  is given by 
\begin{align}
W_2^2 &= \int_0^\tau d\bz \frac{\vartheta_1(\bz - \frac{1}{2} - \frac{L}{2} + \frac{kL}{N}  )}{\vartheta_1(\bz - \tfrac{1}{2} + \tfrac{L}{2})^{k/n}\vartheta_1(\bz - \tfrac{1}{2} - \tfrac{L}{2})^{1-k/n}}.
\end{align}
The high temperature expansion of the above integral (in powers of $e^{-2\pi T}$) is given in Appendix \ref{integral} in equation \eqref{hightemp2}. We can then do the sum over $k$, take the logarithm and take the limit $n\rightarrow 1$ to obtain the following quantities
\begin{align}\label{w2224}
 \sum_{k=1}^{n-1}  \log \Big|\frac{W_2^2 (k,n)}{\beta}\Big| &=  \frac{\pi(n^2-1)L^2}{6\beta} 
\nn \\ & + \left[2\left(  n - \frac{\sinh^2(\pi L/\beta )}{n\sinh^2(\pi L/(n\beta))}   \right) + \frac{2(n^2 -1)}{3n}\sinh^2 (\pi L/\beta)\right]e^{-2\pi/\beta} \nn \\ & \hspace*{7cm}+ \mathcal{O}(e^{-4\pi /\beta }) ,  \\ \label{w2225}
\pd _n  \sum_{k=1}^{n-1}  \log \Big| \frac{W_2^2 (k,n)}{\beta}\Big| &=  \frac{\pi L^2}{3\beta} + \left[\frac{10}{3}+ \frac{2}{3} \cosh (2 \pi  L/\beta)-4 \pi  LT \coth (\pi  L/\beta)  \right] e^{-2\pi /\beta  } \nn \\ & \hspace*{7cm}+ \mathcal{O}(e^{-4\pi /\beta  }). 
\end{align}

\subsection{Bosons}

In this sub-section we put together the high temperature expansions of 
$S_1^{(n)}$ and $S_2^{(n)}$ to obtain the 
R\'{e}nyi  entropies.  
From \eqref{s1} and (\ref{s11}) we have the following\footnote{So far we had been working in units where the length of the spatial circle $R$ is set to unity. We shall now  restore $R$ by $L\rightarrow L/R$ and $\beta \rightarrow \beta/R$.}
\begin{align}
S^{(n)}_1 &=  \frac{n+1}{3n} \log \Big{|}   \frac{R}{\pi T} e^{-\pi  T L^2 / R} \sinh \left( {\pi L T}  \right) \Big|
   -\frac{4(n+1)}{3} e^{-2\pi TR} \sinh^2 \left(  {\pi L T}  \right) \nn \\  & \hspace{9cm}+ \mathcal{O}(e^{-4\pi TR}) ,\\
S_1^{(1)}   &=  \frac{2}{3} \log \Big{|}   \frac{R}{\pi T} e^{-\pi  T L^2 / R} \sinh \left( {\pi L T}  \right) \Big|
   -\frac{8}{3} e^{-2\pi TR} \sinh^2 \left(  {\pi L T}  \right) + \mathcal{O}(e^{-4\pi TR}).
\end{align}
Substituting \eqref{w2224} and (\ref{w2225}) in (\ref{s21}) and  \eqref{s2n}  we get 
\begin{align} 
S_2^{(n)}  &=  \frac{\pi(n+1)L^2}{3\beta} +2\left[ \frac{2}{n-1}\left(  n - \frac{\sinh^2(\pi LT)}{n\sinh^2(\pi LT/n)}   \right) + \frac{2(n+1)}{3n}\sinh^2 (\pi LT)\right]e^{-2\pi TR} \nn \\ & \hspace*{7cm}+ \mathcal{O}(e^{-4\pi TR }) , \\
S_2^{(1)}   &=  \frac{2\pi T L^2}{3R} + 2\left[\frac{10}{3}+ \frac{2}{3} \cosh (2 \pi  LT)-4 \pi  LT \coth (\pi  LT)  \right] e^{-2\pi TR} + \mathcal{O}(e^{-4\pi TR })  \nn \\
&=  \frac{2\pi T L^2}{3R} + \frac{8}{3} \sinh ^2 (\pi LT)  e^{-2\pi TR}  + 8 \left[ 1- \pi LT \coth (\pi LT) \right] e^{-2\pi TR} + \mathcal{O}(e^{-4\pi TR }) .
\end{align}
By adding the two parts above we get the R\'{e}nyi/entanglement entropy to be
\begin{align}\label{boson-re}
S_n =\, &  \frac{n+1}{3n} \log \Big{|}   \frac{R}{\pi T}   \sinh \left( {\pi L T}  \right) \Big| + \frac{4}{n-1} \left[   n - \frac{\sinh^2 (\pi LT  )}{n\sinh^2( \pi LT/n)}   \right] e^{-2\pi RT } + \mathcal{O}(e^{-4\pi T R}) ,\\ 
S_E  =\, &  \frac{2}{3} \log \Big{|}   \frac{R}{\pi T}  \sinh \left( {\pi L T}  \right) \Big|     +  8 \left[ 1- \pi LT \coth (\pi LT) \right] e^{-2\pi TR} + \mathcal{O}(e^{-4\pi TR }).
\end{align}
One can keep track of terms upto any arbitrary order in $e^{-2\pi T R}$. The expression for the entanglement entropy upto $\mathcal{O}(e^{-6\pi T R})$ is 
\begin{align}\label{boson-ee}
S_E  =\, &  \frac{2}{3} \log \Big{|}   \frac{R}{\pi T}  \sinh \left( {\pi L T}  \right) \Big|     +  8 \left[ 1- \pi LT \coth (\pi LT) \right] e^{-2\pi TR}  \nn \\
& +\Bigg{(}  \frac{49}{5} + \frac{14}{15} \cosh (2 \pi  L T)+\frac{4}{15} \cosh (4 \pi  L T)-\frac{21}{2} \pi  L T \text{csch}(\pi  L T) \text{sech}(\pi  L T) \nn \\
& \qquad -\frac{1}{2} \text{sech}^2(\pi  L T)-6 \pi  L T \sinh (2 \pi  L T) \text{sech}^2(\pi  L T) \nn \\
& \qquad +\frac{1}{2} \pi  L T \tanh (\pi  L T) \text{sech}^2(\pi  L T) \Bigg{)} e^{-4\pi TR} + \mathcal{O}(e^{-6\pi TR}).
\end{align}
The low temperature expansion of the R\'{e}nyi and entanglement entropies can 
be obtained directly starting from the low temperature expansions of the 
twist correlators and the integrals of the cut differentials or by the following 
replacements in (\ref{boson-re}) and (\ref{boson-ee})
\begin{equation}\label{replace}
 R\rightarrow \frac{i}{T}, \qquad T\rightarrow -\frac{i}{R} .
\end{equation}
These replacements implement the modular transformation relating the 
low and high temperature expansions. 
The derivative of the \Re entropy in the limit $n\rightarrow 1$ is 
universal in a zero temperature CFT  and is proportional
to the central charge  \cite{Perlmutter:2013gua}. 
However we see that for a finite temperature finite size system of free bosons the derivative 
of \Re entropy is given by 
\begin{align}\label{q4}
S'_n \Big|_{n\rightarrow 1}=&-\frac{1}{3}\log  \Big| \frac{R}{ \pi  T} \sinh  (\pi  L T) \Big| \nn \\
& -4 e^{-2 \pi  TR} \Big(1+2 \pi ^2 L^2 T^2+\pi  L T \left(3 \pi  L T \text{csch}^2(\pi  L T)-4 \coth (\pi  L T)\right)\Big) \nn \\
& +\mathcal{O}(e^{-4\pi TR }).
\end{align}
Therefore there are  finite size corrections to this universal property of the \Re entropy\footnote{We thank Aninda Sinha for raising this point.}.  
   


%

\subsection{Fermions}

We would like to consider the entanglement entropy of fermions on a circle and then check whether the finite size corrections can be 
reproduced by one-loop  calculations in the bulk. 
The entanglement and \Re entropies were calculated in \cite{Azeyanagi:2007bj, Herzog:2013py}. 
We quote the answers here. 
The high temperature expansion of the \Re entropy is given by
\begin{align}
S^{(n)}&=S^{(n)}_1 +S^{(n)}_2  \\
S^{(n)}_1 &= \frac{n+1}{6n}  \log \Big| \frac{\beta}{\pi}e^{-\pi L^2 / \beta} \sinh \left( \frac{\pi L}{\beta} \right) \prod_{m=1}^\infty \frac{ (1-e^{2\pi L/\beta} q^m) (1-e^{-2\pi L/\beta} q^m)   }{(1-q^m)^2} \Big| \\
S^{(n)}_2 &= \frac{n+1}{6n} \frac{\pi L^2}{\beta} - \frac{2}{1-n} \sum_{j=1}^\infty \frac{(-1)^{\nu j}}{j} \frac{1}{\sinh(\pi j/\beta)} 
\left( \frac{\sinh(\pi j L)}{\sinh(\pi j L / (n\beta))} - n \right)
\end{align}
where $\nu (=1,2,3,4)$ denotes the spin structure. We shall work with $\nu=2$ or  (R,NS) fermions, which has the boundary conditions 
\begin{equation}
\psi(z+1)=\psi(z) \ , \qquad \psi(z+\tau)=-\psi(z) \nn 
\end{equation}
To a first few leading orders the \Re entropy is 
\begin{align}\label{fermion-re}
S^{(n)} =  &\frac{n+1}{6n} \log \Big| \frac{R}{\pi T}  \sinh \left( \frac{\pi L}{\beta} \right) \Big|  -  \frac{2(n+1)}{3n}  e^{-2\pi RT} \sinh^2 (\pi L T)  \nn \\ & - \frac{4}{1-n} \left( \frac{\sinh(\pi  L)}{\sinh(\pi  LT / n)} - n \right) e^{-\pi RT} - \frac{2}{1-n}\left( \frac{\sinh(2\pi  L)}{\sinh(2\pi  LT / n)} - n \right) e^{-2\pi RT} \nn \\ &+ \mathcal{O}(e^{-3\pi RT})
\end{align}
The EE given as the sum $S_E = S_1 + S_2$. 
\begin{align}
S_1= \frac{1}{3} \log \Big| \frac{\beta}{\pi}e^{-\pi L^2 / \beta} \sinh \left( \frac{\pi L}{\beta} \right) \prod_{m=1}^\infty \frac{ (1-e^{2\pi L/\beta} q^m) (1-e^{-2\pi L/\beta} q^m)   }{(1-q^m)^2} \Big|
\end{align}
here, $q=e^{2\pi/\beta}$. The leading order behaviour at high-temperatures is then (the circumference of the circle $R$ is restored) 
\begin{align}
S_1 = \frac{1}{3} \log \Big| \frac{R}{\pi T}  \sinh \left( \frac{\pi L}{\beta} \right) \Big| - \frac{\pi T L^2}{3R} - \frac{4}{3} e^{-2\pi RT} \sinh^2 (\pi L T) + \mathcal{O}(e^{-3\pi R T})
\end{align}
The expression for $S_2$ is as follows
\begin{align}
S_2 &= \frac{\pi T L^2}{3R}   - 2\sum_{i=1}^\infty \frac{1}{l} \left[ 1-\frac{\pi L l}{\beta} \coth \left(\frac{\pi L l}{\beta} \right)  \right] \frac{1}{\sinh \left( \frac{\pi l}{\beta} \right)}
\end{align}
The leading order behaviour of $S_2$ is
\begin{align}
S_2 &= \frac{\pi T L^2}{3R} - 4 \left[ 1 - \pi L T \coth \left( \pi L T   \right) \right] e^{-\pi RT} - 2\left[ 1 - 2\pi L T \coth \left(2 \pi L T   \right) \right] e^{-2\pi RT} \nn \\ & \hspace{10.5cm}+ \mathcal{O}(e^{-3\pi RT})
\end{align}
Thus, the entanglement entropy at the first few leading orders at high-temperatures is 
\begin{align}\label{fermion-ee}
S_E =  &\frac{1}{3} \log \Big| \frac{R}{\pi T}  \sinh \left( \frac{\pi L}{\beta} \right) \Big|  - 4 \left[ 1 - \pi L T \coth \left( \pi L T   \right) \right] e^{-\pi RT} \nn \\ & - 2 \left[ 1 - 2\pi L T \coth \left(2 \pi L T   \right) \right] e^{-2\pi RT} - \frac{4}{3} e^{-2\pi RT} \sinh^2 (\pi L T) + \mathcal{O}(e^{-3\pi RT})
\end{align}
Note that just as in the case of bosons, the low temperature expansions for the 
fermions can be obtained by the replacements given in 
equation (\ref{replace}) in the expressions for the R\'{e}nyi entropies and entanglement 
entropies (\ref{fermion-re}) and (\ref{fermion-ee}).  

In the next section we shall show that the leading order one-loop corrections calculated from the bulk exactly reproduce these finite size corrections in \eqref{boson-re}, \eqref{boson-ee}, \eqref{fermion-re} and \eqref{fermion-ee}.

\section{Finite size corrections from holography}
\def\ol{\text{one-loop}}

As mentioned before, in order to  evaluate the \Re entropies we use the replica trick which involves finding the partition function $Z_n$ on the $n$-sheeted cover or a Riemann surface $\Sigma=\mathbb{C}/\Gamma$ with genus $g$. In the dual gravitational theory, we need to evaluate the partition function on a quotient $AdS_3/\Gamma$ which has $\Sigma$ as its conformal boundary \cite{Faulkner:2013yia, Lewkowycz:2013nqa, Barrella:2013wja}. It was shown in \cite{Barrella:2013wja} that in order to obtain the one-loop contribution from the bulk we need to find the Schottky uniformization of the corresponding branched cover ($\Sigma=\mathbb{C}/\Gamma$) and then calculate the one-loop partition function in the bulk quotient $AdS_3/\Gamma$. In this section we shall first review the setup for the computation developed in \cite{Barrella:2013wja} and then utilize it to calculate the one-loop corrections to the entanglement entropy for bosons and  fermions from the bulk. 

\subsection*{Holographic computation of \Re entropies for one interval on a torus }

Since we are interested in a system at finite temperature we require to Euclideanize the temporal direction with a period having the inverse temperature. The spatial direction is also periodic. This gives a torus with two different periodicities. If $z$ is the coordinate on the torus we then have the following doubly-periodic identifications
\begin{align}
z \rightarrow z + R \mathbb{Z} + \frac{i \mathbb{Z}}{T}.
\end{align}
Note that this identification is invariant under the exchange 
\begin{align}
R \rightarrow \frac{i}{T} \ , \qquad T \rightarrow -\frac{i}{R} \nn .
\end{align}
This property turns out to be useful while obtaining high-temperature expansions from low-temperature ones (and vice-versa) of modular invariant quantities. 

Our first task is  to find the discrete group $\Gamma$ by which we can form $AdS_3/\Gamma$ which has $\Sigma=\mathbb{C}/\Gamma$ as its conformal boundary.  Every compact Riemann surface ($\Sigma$) can be obtained as $\mathbb{C}/\Gamma$ with $\Gamma$ being the Schottky group.  The Schottky group  is a discrete sub-group of $PSL(2,\mathbb{C})$. For of a genus $g$ Riemann surface, it is generated by $g$ loxodromic generators, ${L_i}$ where $i=1,2,\cdots,g$. 
It was shown in \cite{Barrella:2013wja} that for the case of the torus the differential equation for Schottky uniformization is
\begin{align}\label{tde}
\psi '' (z) + \frac{1}{2} \sum_{i=1}^2 \left( \Delta \wp (z-z_i) + \gamma (-1)^{i+1} \zeta(z-z_i) + \delta   	\right) \psi(z) =0 .
\end{align}
where $\gamma$ is the accessory parameter, $\Delta =\tfrac{1}{2} (1- \tfrac{1}{n^2})$, $\wp$ is the Weierstrass elliptic function and $\zeta$ is the Weierstrass zeta function. The generators of the Schottky group can be obtained by solving the monodromy problem of the solutions $\psi$ of the above equation. This will be outlined below. 

The derivative of the entanglement entropy is given in terms of accessory parameter as  
\begin{align}\label{ee-der}
\frac{\pd S_E}{\pd z_i} = - \lim _{n\rightarrow 1} \frac{cn}{6(n-1)} \gamma_i .
\end{align}

\subsubsection*{Classical result}

In order to obtain the classical contribution we need to find the accessory parameter in the torus differential equation (\ref{tde}) and then use (\ref{ee-der}). Expanding in $\epsilon(\equiv n-1)$ 
\begin{align}
\psi(z) = \psi_\zero + \epsilon \psi_\one \ , \ \ \Delta = \epsilon \ , \ \ \gamma = \epsilon \gamma_\one \ ,\ \  \delta = - (\pi T)^2 + \epsilon \delta_\one . 
\end{align}
The zeroth and first order solutions are obtained as
\begin{align}
\psi_\zero (z) &= A e^{z\pi T} + B e^{-z\pi T}, \\
\psi_\one (z) &= \frac{e^{-z\pi T}}{2\pi T} \int_0^z e^{x \pi T} m(x) \psi_\zero (x) dx - \frac{e^{z \pi T}}{2\pi T} \int_0^z e^{-x\pi T} m(x) \psi_\zero  (x) dx , 
\end{align}
where
\begin{align}
m(z) =  \frac{1}{2} \sum_{i=1}^2 \left( \Delta \wp (z-z_i) + \gamma_\one (-1)^{i+1} \zeta(z-z_i) + \delta_\one 	\right) . 
\end{align}
By demanding that the first order solution has trivial monodromy around the time circle, one can find the value of the accessory parameter to be
\begin{align}
\gamma_\one = 2\pi T \coth \pi T (z_2 - z_1)  .
\end{align}
Using (\ref{ee-der}) we get the entanglement entropy
\begin{align}
S_E = \frac{c}{6} \log \Big|\sinh ^2 \pi T (z_2 - z_1) \Big|  + \text{const.} 
\end{align}

\subsubsection*{One-loop corrections}

One-loop determinants on quotients of $AdS_3$ have been obtained in \cite{Giombi:2008vd}. The steps followed in order to obtain the one-loop corrections to the entanglement entropy is as follows. 
\begin{enumerate}
\item The Schottky group $\Gamma$ corresponding to $n$-sheeted cover is found. This is done by solving the monodromy problem and then finding the loxodromic generators $L_i$ of the group. 
\item The set of representatives of the primitive conjugacy classes ($\gamma \in \mathcal{P}$) of $\Gamma$ are then found by forming non-repeated words from the $L_i$ and their inverses upto conjugation in $\Gamma$. 
\item Largest eigenvalues of the words ($q_\gamma$) in each primitive conjugacy class $\gamma$ are then computed and substituted into the expressions for the one-loop determinants. We then sum over all primitive conjugacy classes
\begin{equation}\label{conj-sum}
\log Z[n] \Big|_{\text{one-loop}} = \sum_{\gamma \in \mathcal{P}}  \log  Z(q_\gamma) . 
\end{equation}
where $Z(q)$ is the one loop determinant  of the 
bulk fields in thermal $AdS_3$ with modular parameter $q$. 
\item The one-loop correction to the \Re entropy can found using
\begin{align}
S_n \Big|_{\ol}= \frac{1}{1-n} \left( \log Z[n]\Big|_{\ol} - n \log Z[1]\Big|_{\ol} \right) .
\end{align}
The correction to the entanglement entropy is then obtained by taking the $n\rightarrow 1$ of the above answer. 
\end{enumerate}

We shall choose the interval as $[-y,y]$ i.e.\,\,the length of the interval is $L=2y$. We shall calculate the one-loop correction in the high-temperature limit. 
We define a new coordinate $u$ as
\begin{align}
u \equiv e^{-2\pi T z}
\end{align}
Defining $u_y = e^{2\pi T y }$, $u_R = e^{2\pi T R}$ and $f(u)=u+u^{-1}$ we can write the Weierstrass elliptic function as
\begin{align}
\wp (z \pm y) &= \sum_{m=-\infty}^{\infty} \frac{4\pi^2 T^2}{f(uu_y^{\pm 1} u_R^m)} - \sum_{m\neq 0} \frac{4\pi^2 T^2}{f(u_R^m)} + \frac{\pi^2 T^2}{3}, 
\end{align}
and a similar expression for $\zeta(z\pm y)$. 

The torus differential equation is then solved for one spatial period $-R/2 \geq \text{Re} \ z \geq R/2$. The ansatz for the solutions is taken as
\begin{align}
\psi^\pm &= \frac{1}{\sqrt{u}} (u - u_y)^{\Delta_\pm} \left( u - \frac{1}{u_y} \right)^{\Delta_\mp} \sum_{m=-\infty}^\infty \psi^{\pm (m)} (u_y, u_R) u^m , 
\end{align}
where $\Delta_\pm= \frac{1}{2}(1\pm \frac{1}{n})$ and the solutions are normalized as $\psi^{\pm(0)}=1$. The coefficients $\psi^{\pm (m)}$ are then expanded in a power series in $u_R$ in the following form
\begin{align}
\psi^{\pm(m)} (u_y, u_R) = \sum_{k=|m|}^\infty \psi^{\pm(m,k)} (u_y) u_R^k . 
\end{align}
The differential equation (\ref{tde}) is then expanded in $u$ and $u_R$. The coefficients $\psi^{\pm (m,k)}$ can then be determined. Also one can obtain $\gamma$ and $\delta$ upto any order in $u_R$
\begin{align}
\gamma &= \pi T (1+u_y^2) \left[ \frac{1-n^2}{n^2(u_y^2 -1 )} + \frac{(n^2 -1 )^2 (u_y^2 -1 )^3}{6n^4u_y^4} u_R^2 + \mathcal{O}(u_R^4) \right] , \\
\delta &= \frac{\pi^2 T^2}{n^2(u_y^2 -1)} \Big{\lbrace} \frac{1}{6} \left[ (n^2 -1)(u_y^2 + 1) \log (u_y) - (7n^2 -1) (u_y^2 -1) \right] + \mathcal{O}(u_R^2) \Big{\rbrace} . 
\end{align}
The classical \Re entropy can then be obtained using (\ref{ee-der})
\begin{align}\label{classical}
S_n = \frac{c(n+1)}{12n} \left[ \log \sinh ^2 (2\pi T y) + \text{const.} - \frac{(n^2 -1)}{6n^2} \left[ \cosh (8\pi T y) - 4 \cosh (4 \pi T y) \right] e^{-4\pi T R} \right] . 
\end{align}
The solutions to the first few orders in $u_R$ is
\begin{align}
&\psi ^+ (u) = \frac{1}{\sqrt[]{u}} (u - u_y)^{\Delta_+} \left( u - \frac{1}{u_y} \right)^{\Delta_-} \Bigg[ 1 + \frac{(n^2 -1 )(u_y^2 -1)^2 u_R^2}{24n^3 u^2 u_y^3}  \times  \nn \\
& \ u((n+1)u^2 +n -1)u_y^2 + u((n-1)u^2 +n +1)-n(u^4+1)u_y + \mathcal{O}(u_R^3)                  \Bigg] . 
\end{align}
The solution for $\psi^-$ is same as the above with $u\rightarrow 1/u$. The monodromy matrices, $L_1$ can then be found by
\begin{align}
\begin{pmatrix}
\psi^+(u/u_R) \\
\psi^- (u/u_R) 
\end{pmatrix} = L_1 \begin{pmatrix}
\psi^+ (u) \\
\psi^- (u) 
\end{pmatrix} . 
\end{align}
To the lowest few orders we get matrix elements to be
\begin{align}
(L_1)_{11}&= \frac{n\uy^{1-1/n}}{(1-\uy ^2)\sqrt{\ur}} \left\lbrace  	1 - \frac{[(n+1)\uy ^2 +n -1]^2}{4n^2 \uy^2}	+\o (\ur^2)	  \right\rbrace \nn \\
(L_1)_{11}&= \frac{n\uy}{(1-\uy ^2)\sqrt{\ur}} \left\lbrace  	1 - \frac{[(n+1)\uy ^2 +n -1][(n+1)\uy ^2 +n +1]}{4n^2 \uy^2}	+\o (\ur^2)	  \right\rbrace \nn \\
(L_1)_{21}&=-(L_1)_{12} \ , \qquad (L_1)_{22}= (L_1)_{11}|_{n\rightarrow n } . 
\end{align}
The other Schottky generators $L_i$, $i=2,3,\cdots, n$ can be obtained by conjugating $L_1$ by $M_2$ around $z_2=y$
\begin{align}
L_i = M_2^{i-1} L_1 M_2^{-(i-1)} \ , \qquad M_2 = \begin{pmatrix}
e^{2\pi i \Delta_+} &0 \\
0 & e^{2\pi i \Delta_-} 
\end{pmatrix}. 
\end{align}
Words of length $k$ can then be constructed as follows
\begin{align}
&L_{k_1}^{\sigma_1} L_{k_2}^{\sigma_2} \cdots L_{k_m}^{\sigma_m} = \left[ \frac{nu_y}{(1-u_y^2)\sqrt{u_R}} \right]^m 
\left[ \prod_{j=1}^{m-1} \sigma_j \left( u_y^{-\sigma_j/n} - e^{2\pi i (k_j - k_{j+1})/n} u_y^{\sigma_{j+1}/n} \right) \right]  \nn \\
& \quad \times \begin{pmatrix}
\sigma_m u_y^{-\sigma_m /n} 	&\sigma_m e^{2\pi i k_m/n} \\
-\sigma_m e^{-2\pi i k_1 /m} u_y^{(\sigma_1 - \sigma_m)/n} &-\sigma_m e^{2\pi i (k_m -k_1)/n} u_y^{\sigma_{\sigma_1/n}}
\end{pmatrix} + \mathcal{O}(u_R^{-m/2+1}) .
\end{align}
here $j=1,2,\cdots,m$ and $\sigma_j=\pm 1$. The larger eigenvalue of the above word can then be found out to be
\begin{align}\label{q}
q_\gamma^{-1/2} = \left[ \frac{nu_y}{(1-u_y^2)\sqrt{u_R}} \right]^m \prod_{j=1}^\infty \sigma_j \left( u_y^{-\sigma_j/n} - e^{2\pi i (k_j - k_{j+1})/n}u_y^{\sigma_{j+1}/n}  \right) +\mathcal{O}(u_R^{-m/2+1}) .
\end{align}
here $m+1\rightarrow 1$.
One can then substitute the above expression into the formulae for \ol \ partition functions in 3-d gravity and then sum over primitive conjugacy classes (i.e.\,\,over all possible non-repeated words) \eqref{conj-sum}. The corrections to the entanglement entropy can be obtained by using
\begin{align}\label{ee-ol}
S_E|_\ol=\lim_{n\rightarrow 1}S_n|_\ol =\lim_{n\rightarrow 1} - \frac{1}{n-1}\Big( \log {Z[n]}|_\ol - \log {Z[1]}|_\ol      \Big) . 
\end{align}

\subsection{One-loop corrections: Bosons}\label{sec1loopb}

In the CFT we used the replica trick to evaluate the R\'{e}nyi entropies. 
This essentially is the partition function of the CFT on the $n$-sheeted cover of the 
 CFT on the torus. 
 As we have argued in the introduction we expect the leading 
   finite size corrections to the  R\'{e}nyi entropies of the free boson CFT should  agree 
with the one-loop determinant of the corresponding bulk field evaluated in the 
handlebody geometry dual the $n$-sheeted  cover of the CFT on the torus. 
In this section we will evaluate the one-loop determinant  of bulk fields in the 
handlebody geometry dual to the $n$-sheet cover of the CFT and show that the  leading finite 
size corrections  evaluated in the CFT  agree precisely with that from  the one-loop determinants.

The  free real scalar in 1+1 dimensions is 
not a good primary. The well defined primary with 
dimension $(\Delta, \bar\Delta) = (1, 0)$ or $(0, 1)$ are 
conserved currents constructed from the scalar.
 If $\phi(z, \bz)$ is a real scalar field, the two conserved currents are then given by 
\begin{align}
J(z) = (\pd _z \phi) (z) \ , \qquad J(\bz) =  (\pd _\bz \phi) (\bz)  . 
\end{align}
The conservation law is then guaranteed by the equations of motion : $\pd_\bz \pd_z \phi(z,\bz) =0$. 
The presence of these (anti-)holomorphic currents in the CFT is equivalent to the presence of two  $U(1)$ Chern-Simons gauge fields in the bulk. The one-loop determinant for such a spin-1 field 
in thermal $AdS_3$
can be calculated by using heat kernel methods 
\cite{David:2009xg,justin} and is explicitly shown in Appendix \ref{chern-simons}. 
\begin{align}\label{gauge-det}
Z^\ol_{\text{spin-1}} &=\left( \prod_{n=1}^{\infty} \frac{1}{(1-q^{m})^{1/2}(1-\bar{q}^{m})^{1/2}} \right)^2 .
\end{align}
Note that we have taken the square because the the free boson of interest is complex, therefore
there are $4$ Chern-Simons fields in the bulk. 
Using \eqref{q} we get
\begin{align}
q_1^{-1/2} =& \frac{n\uy(\uy^{-1/n}-\uy^{1/n})}{(1-\uy^2)\sqrt{u_R}} + \mathcal{O}(\sqrt{u_R}) .
\end{align}
Using \eqref{q} and the expression for the one-loop partition function \eqref{gauge-det}, we can obtain the one-loop partition function for the $n$-sheeted cover from \eqref{conj-sum}. We need to sum only  over all single-letter words  to obtain corrections upto the leading order
\begin{align}
\log Z_n |_\ol = \sum_{\gamma \in \mathcal{P}} \text{Re} [2 q_\gamma + \mathcal{O} (q_\gamma^2) ]  = \frac{4 \sinh^2(2\pi T y)}{n\sinh^2 (2\pi T y /n)} e^{-2\pi T R} + \mathcal{O}(e^{-4\pi T R}) .
\end{align}
The one-loop correction to the \Re entropy is 
\begin{align}\label{1loopbos}
S_n = \frac{4}{n-1} \left[   n - \frac{\sinh^2 (\pi T L)}{n\sinh^2( \pi TL/n)}   \right] e^{-2\pi T R} + \mathcal{O}(e^{-4\pi T R}) .
\end{align}
Finally, using \eqref{ee-ol} we can calculate the one-loop correction to the entanglement entropy 
\begin{align}
S_E |_\ol = 8(1- \pi T L \coth (\pi T L) ) e^{-2\pi T R} + \mathcal{O}(e^{-4\pi T R}) . 
\end{align}
(where $L=2y$). 
This matches precisely with the leading order correction to the universal part of the entanglement entropy obtained in \eqref{boson-ee}.

\def\nm{{  \frac{n-1}{2}  }}
\subsection{One loop corrections: Fermions}

In this sub-section we repeat the analysis of the comparison of  one-loop determinants in the 
handlebody geometry to the finite size corrections evaluated in the CFT for the 
case of free fermions.  
 We will see that the leading and the 
next-to leading finite size corrections agree evaluated in the CFT agree with the bulk 
one-loop determinant. Demonstrating this 
involves summing over 
two letter words to evaluate the bulk determinant. 
Since $(i\slashed{\pd})^2 =- \nabla^2_{(1/2)}$, the one-loop partition function for the spin-$\frac{1}{2}$ fermion is 
\begin{align}
Z^\ol_{\text{spin-1/2}} = \det  (i\slashed{\pd}) =\sqrt{ \det  (-\nabla^2_{(1/2)})}  . 
\end{align}
	Determinants of laplacians for fermions in the BTZ black hole were calculated using quasinormal modes and also by integrating the heat-kernel in \cite{Datta:2012gc}. The one-loop partition function 
	of the bulk field dual to operators of conformal dimension 
	$(\Delta, \bar \Delta) =( \frac{1}{2}, 0) $  and $(0, \frac{1}{2})$  reads
\begin{align}
Z^\ol_{\text{spin-1/2}} &= \prod_{n=0}^{\infty} |1-q^{n+1/2}| ^{2(n+1)} . 
\end{align}
The high temperature expansion of the one-loop determinant on the $n$-sheeted cover \eqref{conj-sum}  reads as
\begin{align}\label{ferm-det}
\log Z_n|_\ol &=\sum_{\gamma \in \mathcal{P}} \text{Re} \left[  -2q_\gamma^{1/2} -   q_\gamma + \mathcal{O}(q_\gamma^2)  \right] . 
\end{align}
We have kept the first two leading orders in the high-temperature expansion. At $\mathcal{O}(e^{-\pi R T })$, the contribution is just from the sum over single letter words. However at  $\mathcal{O}(e^{-2\pi R T })$, there is one contribution from the sum over two-letter words from $-2q_\gamma^{1/2}$ and another from the sum over  the squares of the single letter words from $-q_\gamma$.

While summing over 2-letter words ($L^{\sigma_1}_{k_1}L^{\sigma_2}_{k_2}$) we need to keep the following constraints in mind
\begin{enumerate}
\item  Since the sum is over primitive conjugacy classes  (or non-repeated words), we need to impose $k_1 \neq k_2$. Another way to say this is that, we would be summing over squares of single-letter words if $k_1=k_2$ which is already taken care of by terms in one-loop partition function which are higher order in $q_\gamma$.
\item We need to ensure that each $(k_1,k_2)$ pair in the sum is counted only once. This is because $L^{\sigma_1}_{k_1}L^{\sigma_2}_{k_2}$ and $L^{\sigma_2}_{k_2}L^{\sigma_1}_{k_1}$  are simply related by conjugation and hence belong to the same conjugacy class. 
\item Words and their inverses cannot be related by conjugation and need to summed over separately.
\end{enumerate}
 The high-temperature expansion can then be written as (the prime on the $k_2$ sum indicates the constraints above)
\begin{align}\label{fermion-sum}
\log Z_n|_\ol &= -2\sum_{k=0}^{n-1} \text{Re} [q_k^{1/2}] - 2 \sum_{k_1=0}^{n-1} {\sideset{}{'}\sum_{k_2=0}^{n-1}} \sum_{\sigma_1 , \sigma_2} \text{Re} [(q^{\sigma_1,\sigma_2}_{k_1,k_2})^{1/2}] \nn \\
&\qquad\qquad  -  \sum_{k=0}^{n-1} \text{Re} [q_k] + \mathcal{O}(e^{-3\pi RT }) . 
\end{align}

Let us now list eigenvalues of all possible 2-letter words. The 2-letter words are of the form $L_{k_1}^{\sigma_1}L_{k_2}^{\sigma_2}$ with $\sigma_{1,2}=\pm 1$ and $k_{1,2}=0, 1, \cdots, n-1$. 
\begin{align}\label{2l-e}
(q^{\pm \pm}_{k_1,k_2})^{-1/2} &= \frac{n^2}{(u_y^{-1} - u_y)^2 u_R} \left( u_y ^{-1/n} - e^{2\pi i (k_1 -k_2)/n} u_y^{1/n} \right)   \left( u_y ^{-1/n} - e^{-2\pi i (k_1 -k_2)/n} u_y^{1/n} \right)\nn \\ & \hspace{11cm} +\mathcal{O}(1) , \nn \\
(q^{\pm \mp}_{k_1,k_2})^{-1/2} &= -\frac{n^2}{(u_y^{-1} - u_y)^2 u_R}  \left(1 - e^{2\pi i (k_1 -k_2)/n}   \right)   \left( 1 - e^{-2\pi i (k_1 -k_2)/n}   \right) + \mathcal{O}(1) . 
\end{align}
The sums over the two-letter words in  \eqref{fermion-sum}  are done in Appendix \ref{app-sum}.   We then have the following result for the one-loop partition function.  
\begin{align} \label{fermion-Z}
\log Z_n|_\ol =& - 4 \frac{\sinh (2\pi Ty)}{\sinh (2\pi T y/n)} e^{-\pi TR} -2 \left[  \frac{\sinh (4\pi Ty)}{\sinh (4\pi T y/n)} - \frac{\sinh^2 (2\pi Ty)}{n\sinh^2 (2\pi T y/n)}  \right] e^{-2\pi TR}  \nn \\
&+\frac{2(n^2-1)}{3n} \sinh ^2 (2\pi Ty) e^{-2\pi TR} - \frac{2\sinh^2 (2\pi Ty)}{n\sinh^2 (2\pi T y/n)} e^{-2\pi TR}  + \mathcal{O}(e^{-3\pi TR}) . 
\end{align}
The first term on the RHS is the single letter word contribution from the $q_k^{1/2}$ sum. The second and third term arises from the two-letter word sums $(q^{\pm\pm}_{k_1,k_2})^{1/2}$ and $(q^{\pm\mp}_{k_1,k_2})^{1/2}$ respectively. The fourth term is the single letter word sum over $q_k$. Upon simplification, we get 
\begin{align} 
\log Z_n|_\ol =& - 4 \frac{\sinh (2\pi Ty)}{\sinh (2\pi T y/n)} e^{-\pi TR} -2 \frac{\sinh (4\pi Ty)}{\sinh (4\pi T y/n)}   e^{-2\pi TR} \nn \\
&+\frac{2(n^2-1)}{3n} \sinh ^2 (2\pi Ty) e^{-2\pi TR}  + \mathcal{O}(e^{-3\pi RT}) . 
\end{align}
The one-loop correction to the \Re entropy can then be calculated to be 
\begin{align}\label{1loopferm}
S_n |_\ol=& - \frac{1}{n-1} \left[  4\left( n -  \frac{\sinh (2\pi Ty)}{\sinh (2\pi T y/n)} \right)   e^{-\pi RT} 	+2\left( n -  \frac{\sinh (4\pi Ty)}{\sinh (4\pi T y/n)} \right)   e^{-2\pi RT} 			  \right] \nn \\
&  -\frac{2 (n+1)}{3 n} \sinh ^2 (2 \pi  T y)  e^{-2\pi RT} 		 + \mathcal{O}(e^{-3\pi RT}) .
\end{align}
Taking the $n\rightarrow 1$ limit,   we can calculate the one-loop correction to the entanglement entropy 
\begin{align}
S_E |_\ol =& -4[1- \pi T L \coth (\pi T L) ] e^{-\pi  RT}   - 2 \left[ 1 - 2\pi L T \coth \left(2 \pi  TL   \right) \right] e^{-2\pi RT} \nn \\
&- \frac{4}{3} e^{-2\pi RT} \sinh^2 (\pi  TL) + \mathcal{O}(e^{-3\pi RT}) .
\end{align}
Again, these one-loop contributions to the \Re and entanglement entropies agree precisely with the leading order correction to the universal part   for the fermion case \eqref{fermion-re} and \eqref{fermion-ee}.

\subsection{Discussion} \label{discuss}

At the first sight it might seem surprising 
 that the one-loop corrections to the R\'{e}nyi entropies 
 from holography, which usually describes strongly coupled physics,  exactly reproduces the leading order finite size corrections of free bosonic and fermionic CFT's. 
 But as discussed in the introduction,  one loop corrections of bulk fields in the 
 handlebody geometries are sensitive to only the quantum numbers of the spectrum
 in the bulk and are insensitive to interactions. 
 The one loop corrections in the bulk can be obtained just by summing  the contributions
 over the spectra in the bulk. 
 The one loop determinant of the Chern-Simons field  in the handlebody geometries can be 
 obtained by substituting (\ref{gauge-det}) into (\ref{conj-sum}) as we have done in 
subsection \ref{sec1loopb}. 
 Similarly the corresponding equation for the Dirac fermion are given by substituting (\ref{ferm-det})
 into (\ref{conj-sum}).   Note that these one-loop determinants organize into 
 representations of  Virasoro characters. 
 The leading  finite size expansions of the one-loop contributions  given in 
 (\ref{1loopbos})  and (\ref{1loopferm}) 
 for the Chern-Simons field and for the Dirac fermions respectively.   
 The contribution to finite size corrections from the classical part of the action 
 given in (\ref{classical}) 
 is clearly sub-leading  to the leading terms from the one-loop terms. 
 The one loop contribution to the partition function is independent of the central charge or 
 the 3-dimensional Newton's constant and just depends on the quantum numbers of the
 spectrum. Therefore  from the usual rules of AdS/CFT,  as summarized in 
 the equality of the partition functions in equation (\ref{adscft}) we expect the leading 
 terms in (\ref{1loopbos}) for the Chern-Simons field and (\ref{1loopferm}) for the Dirac field to agree 
 with the leading finite size corrections evaluated in the CFT given in (\ref{boson-re}) 
 and (\ref{fermion-re}).
 We have shown in this paper by explicit computation of these correction both from the 
 bulk and in the CFT that this expectation is borne out. 
 This agreement is certainly a non-trivial test of the methods proposed by \cite{Barrella:2013wja} to evaluate
 finite size corrections in the bulk. 
 
 Note that the $1/c$ corrections to the partition functions in the bulk 
 which occur at higher loops are sensitive
 to the OPE's of the dual CFT and certainly finite size corrections from these
 would not in general agree with the free boson or the free fermion CFT
 \cite{Giombi:2008vd}. 
 Another point worth mentioning is that we expect the leading classical  contribution to the R\'{e}nyi entropies from the bulk which is proportional 
 to $c$ for  which few terms were evaluated  
 in (\ref{classical}) to be universal for large $c$ interacting CFT's. 
 The reason is that this term just depends on just on the classical action of the handlebody 
 geometry or in other words the metric of the background. 
 It will be interesting to reproduce the  classical action including the finite size corrections
 from this term  by studying 
 large $c$ CFT's using 
 methods similar to \cite{Hartman:2013mia}. 
 This work  obtained  the universal features of the R\'{e}nyi entropies of 
 two intervals in large $c$ CFT's and showed that it reduces to the method 
 proposed by \cite{Faulkner:2013yia} in the bulk.

\section{Conclusions}

In this paper we have calculated the \Re entropies for the free boson on a circle at finite temperature using two point correlation functions of twist field operators. The answer \eqref{renyi-full} is expressed in terms of Jacobi elliptic theta functions and the Riemann-Siegel function which can be written as the $A_{n-1}$ lattice sum. We have also obtained the entanglement and \Re entropies in the decompactification regime and have numerically plotted them. The behaviour of the entanglement entropy was then analysed in the limit of large intervals (equalling the size of the system) and it was shown to match with the thermal entropy as expected. 
We have then  set up a systematic high temperature expansion for the \Re entropies  and have
 extracted the leading order finite-size corrections to the universal part of the \Re and entanglement entropies in the high temperature expansion. 
  As discussed in the introduction and also in section \ref{discuss}
  these leading corrections from the free boson CFT 
 are expected to agree with 
 the leading finite size contributions to the \Re entropies  from  the one-loop determinant of 
 the Chern-Simons field in the handlebody geometries dual to the CFT on 
 $n$-sheeted Riemann surfaces. 
 These corrections were evaluated from the bulk 
  using methods developed recently in \cite{Barrella:2013wja} and were shown to 
  agree precisely with that from the free boson CFT. 
  We have also investigated the case of fermions in which the \Re and entanglement entropies were calculated previously in \cite{Azeyanagi:2007bj, Herzog:2013py}. 
 We have shown that   the leading  and the next to leading order finite size corrections to the universal piece agree with that obtained from holography.  This involved an agreement up to two letter words  composed out of the Schottky generators. 
 
%

The above analysis of free bosons along with the one for free fermions in \cite{Herzog:2013py} gives results for the entanglement entropy on the torus for two of the most well-studied and simplest conformal field theories in 1+1 dimensions. These results will be  useful  to find entanglement and \Re entropies of  superconformal  field theories in 1+1 dimensions like the D1/D5 system which is 
one of the well studied examples of $AdS_3/CFT_2$. 
In fact it will be interesting to figure out how the dependence on the size of the
target space torus $\R$ can be seen in the bulk.

These results also serve as a test for the holographic methods 
using Schottky uniformization developed in \cite{Faulkner:2013yia, Barrella:2013wja}  for the case of finite temperature CFT's with a compact spatial direction.  
The calculation of R\'{e}nyi entropies require the evaluation of partition functions of CFT's on higher genus surfaces. 
Therefore, as emphasized in the introduction our results also test the proposals 
of \cite{Giombi:2008vd,Yin:2007gv}
for the evaluation of one-loop contributions to partition functions 
of bulk fields in handlebody geometries. 

The  result for R\'{e}nyi entropies for the  free boson CFT
 can also be applied to a wide class of one-dimensional statistical models at criticality which are described by $c=1$ or bosonic CFT's. 
Examples include one-dimensional Bose gases, Heisenberg spin chains, the 4-state Potts model and Z$_4$ parafermions.  In this regard 
the dependence on the size of the spacetime torus $\R$ will play an important role.
These results might also serve as approximations in more realistic systems of bosons in 1-dimension  at the critical point like  trapped cold atoms,  superfluids in nanopores  and superconducting wires \cite{Rigol}. These systems are of both theoretical and experimental interest in recent times.  

We end with some future directions and some open questions. One can hope to generalize the results of this work for the case of two or more disjoint intervals and then extract meaningful quantities like the mutual information.  In fact  correlators involving arbitrary insertions 
of bosonic twist fields on the torus have been evaluated in \cite{Atick:1987kd}. 
Another useful direction is to obtain exact results for the entanglement entropy for the free boson in presence of chemical potentials along the similar lines of what is done for the case of fermions \cite{Herzog:2013py}. Holographic studies of the  situation 
with chemical potentials and generalization 
to CFT's  in higher dimensions is also an interesting direction to explore  \cite{Caputa:2013eka,Belin:2013uta}.  

\section*{Acknowledgements}

We are grateful to  Rajesh Gopakumar for stimulating discussions and  useful suggestions. 
We also thank Sachin Jain, Shaon Sahoo, Ashoke Sen  and Aninda Sinha
for useful  discussions.  We thank Sandeep Chatterjee and Abhishek Iyer for help with the
numerics. J.R.D thanks  the string theory group at
 Harish-Chandra Research  Institute, Allahabad for hospitality 
during which part of the work was done, an 
 opportuntity to present this work and for the stimulating discussions 
during the presentation.  The work of J.R.D is partially supported by the 
Ramanujan fellowship DST-SR/S2/RJN-59/2009. 

\def\vt{\vartheta}
\appendix
\section*{Appendix}

\section{Jacobi theta functions}

In this appendix we shall list some identites involving Jacobi theta functions which are used throughout the paper.
The basic theta functions are defined by the series
\begin{align} \label{theta-def}
\vt (z|\tau) = \sum_{n=-\infty} ^\infty \exp \left( \pi i n^2 \tau + 2\pi i n z  \right).
\end{align}
The product representations for theta functions with different characteristics or spin structures are :
\begin{align}
\vt_1 (z|\tau) &= 2 e^{\pi i \tau /4} \sin(\pi z) \prod_{m=1}^\infty (1-q^m)(1-yq^m)(1-y^{-1}q^m) ,\\
\vt_2 (z|\tau) &= 2 e^{\pi i \tau /4} \cos(\pi z) \prod_{m=1}^\infty (1-q^m)(1+yq^m)(1+y^{-1}q^m) , \\
\vt_3 (z|\tau) &=  \prod_{m=1}^\infty (1-q^m)(1+yq^m)(1+y^{-1}q^m) , \\
\vt_4 (z|\tau) &=  \prod_{m=1}^\infty (1-q^m)(1-yq^m)(1-y^{-1}q^m)  ,
\end{align}
where $q=e^{2\pi i \tau}$ and $y=e^{2\pi i z}$ is used\footnote{Note that   {Mathematica} uses the older convention for the nome, $q=e^{\pi i \tau}$ in \texttt{EllipticTheta}.}.\\
The functions have the following (quasi-)periodicity properties
\begin{align}
\vt_\nu (z+1|\tau) &= \vt_\nu (z|\tau), \\
\vt_\nu (z+\tau | \tau) &= e^{-\pi i \tau - 2\pi i z}\, \vt_\nu (z|\tau)  \qquad \text{for }\nu = 1,2,3,4.
\end{align}
The S-modular transformation identities are 
\begin{align}
\vt_1 (z|\tau) &= - i(-i\tau)^{-1/2} e^{-\pi i z^2 /\tau } \, \vt_1 \left(\frac{z}{\tau}\Big|-\frac{1}{\tau}\right), \\
\vt_2 (z|\tau) &=  (-i\tau)^{-1/2} e^{-\pi i z^2 /\tau } \, \vt_4 \left(\frac{z}{\tau}\Big|-\frac{1}{\tau}\right) ,\\
\vt_3 (z|\tau) &=  (-i\tau)^{-1/2} e^{-\pi i z^2 /\tau } \, \vt_3 \left(\frac{z}{\tau}\Big|-\frac{1}{\tau}\right), \\
\vt_4 (z|\tau) &=  (-i\tau)^{-1/2} e^{-\pi i z^2 /\tau } \, \vt_2 \left(\frac{z}{\tau}\Big|-\frac{1}{\tau}\right) .
\end{align}
These functions also have the following half-periodicity properties
\begin{align}
\vt_2(z|\tau) &= - \vt_1 (z-1/2|\tau) ,\\
\vt_3(z|\tau) &= - y^{-1/2}q^{1/8}\, \vt_1 (z-1/2-\tau/2 |\tau), \\
\vt_4(z|\tau) &= i y ^{-1/2} y^{1/8}\, \vt_1 (z-\tau/2|\tau) .
\end{align}
The Dedekind eta function is 
\begin{align}
\eta(\tau) = e^{\pi i \tau /12 }  \prod_{m=1}^\infty ( 1- q^m) .
\end{align}

\section{Evaluating $W_1^1$ and $W_2^2$}
\label{integral}

\subsection*{Reality properties}

In this subsection we make the following claims regarding the elements of the matrix $W$ and prove them. 
\begin{align}\label{reality}
(W_1^1)^* = W_1^2 ,\ \quad (W_1^1)^* = W_1^1 ,\ \quad (W_2^1)^* = -W_2^1 ,\ \quad (W_2^1)^* =  W_2^2 .
\end{align}
These identities imply 
\begin{itemize}
\item \(W_1^1\) and \(W_1^2\) are purely real. 
\item \(W_2^1\) and \(W_2^2\) are purely imaginary.
\item \( \det (W)\) and \( \det(\bar{W})\) are purely imaginary. $\det(W) = 2i \,\im (W_1^1 W_2^2) $. 
\end{itemize}

\begin{figure}[!t]
\centering
\begin{tabular}{c}
\includegraphics[width=3in]{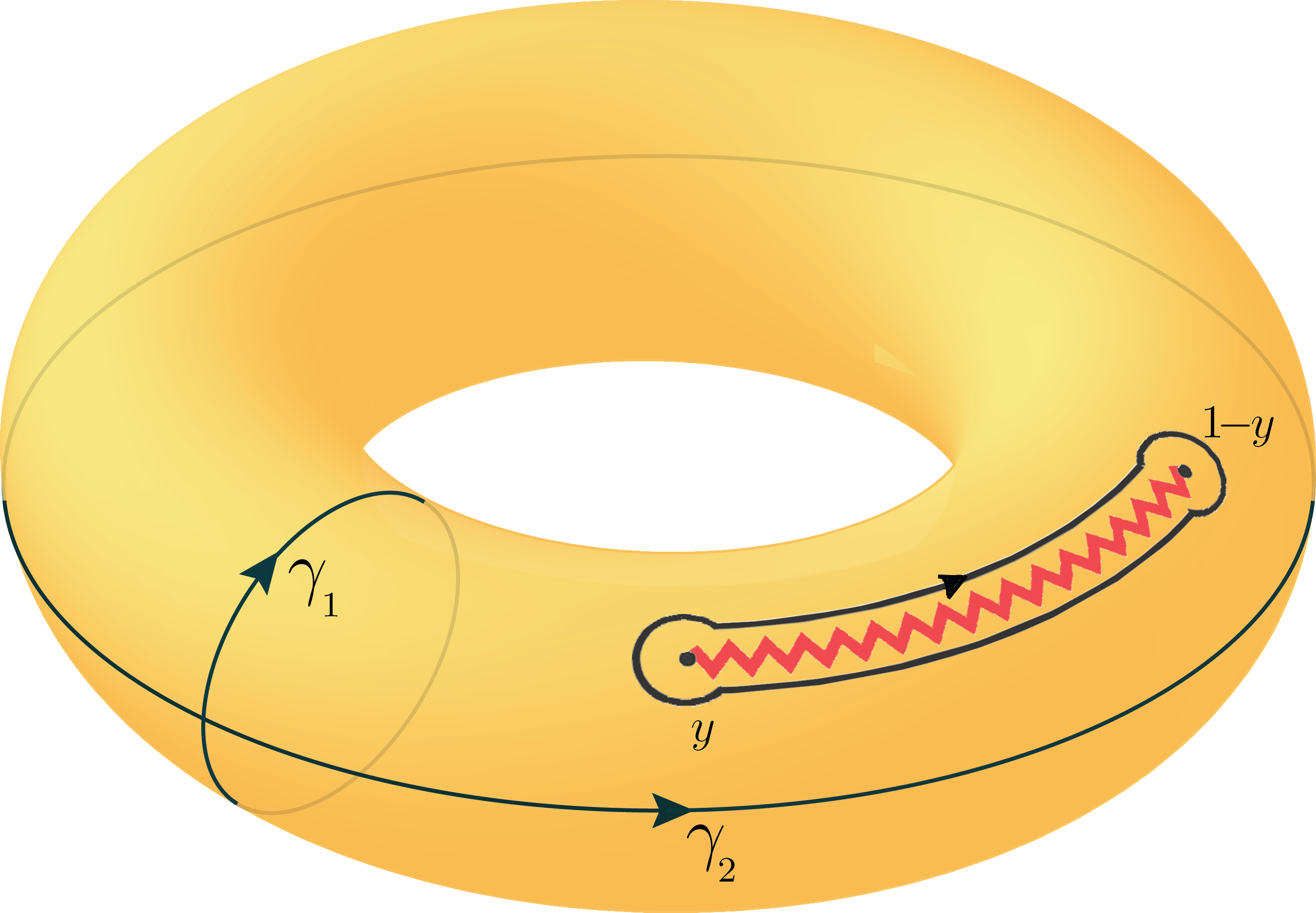} 
\end{tabular}
\caption{\small  The deformable contour on the  torus. }
\label{torus-contour}
\end{figure}

We now proceed to a proof of (\ref{reality}). We list the explicit integral forms of $W_i^j$ on choosing $z_1=y$ and $z_2=1-y$ with $0\leq y \leq \frac{1}{2}$. We also define $y_M = (1-\tfrac{2k}{n})y + \tfrac{k}{n} $. 
\begin{align}\label{W-int}
W_1^1 (k,n)&= \int _0^1 dz \,  \vartheta_1(z-y)^{-(1-k/n)} \vartheta_1(z-1+y)^{-k/n} \vartheta_1 (z - y_M ), \\
W_1^2 (k,n)&= \int_0^1 d\bz \, \vartheta_1( z - \by)^{-k/n}\vartheta_1(\bz -1+ y)^{-(1-k/n)} \vartheta_1 (\bz + y_M ) ,\\
W_2^1 (k,n)&= \int _0^\tau dz \,  \vartheta_1(z-y)^{-(1-k/n)} \vartheta_1(z-1+y)^{-k/n} \vartheta_1 (z - y_M ), \\
W_2^2 (k,n)&= \int_0^\tau d\bz \, \vartheta_1( \bz -  y)^{-k/n}\vartheta_1(\bz-1+  y)^{-(1-k/n)} \vartheta_1 (\bz + y_M ) .
\end{align}
It can then be easily seen that
\begin{align}\label{w01}
\det (W(0,1))= 2i \,\im (W_1^1 (0,1) W_2^2(0,1) ) = 2\tau
\end{align}
Observe that there exists a branch cut from $[y,1-y]$ see Fig.\,\ref{torus-contour}. Let us first calculate the value of the integral along this branch cut. 
\begin{align}
I = \int_y^{1-y} dz \, \vartheta_1(z-y)^{p} \vartheta_1(z-1+y)^{q} \vartheta_1 (z - y_M ),
\end{align}
where, $p$ and $q$ are some rational fractions. Taking a contour `around' this branch cut and calling the contribution to be $J$ we have
\begin{align}\label{residue}
J&=\left(  \int_{y+i\epsilon}^{1-y+i\epsilon} - \int_{1-y-i\epsilon}^{y-i\epsilon} \right) dz \, \vartheta_1(z-y)^{p} \vartheta_1(z-1+y)^{q} \vartheta_1 (z - y_M )\nn \\ &= \text{Sum of residues} \, (=0) 
\end{align}
It can be seen that
\begin{align}
 \int_{y+i\epsilon}^{1-y+i\epsilon} dz \, \vartheta_1(z-y)^{p} \vartheta_1(z-1+y)^{q} \vartheta_1 (z - y_M  ) = e^{iq\pi } I,
\end{align}
and 
\begin{align}
 \int_{1-y-i\epsilon}^{y-i\epsilon}  dz \, \vartheta_1(z-y)^{p} \vartheta_1(z-1+y)^{q} \vartheta_1 (z - y_M  ) = e^{i(2p+q)\pi } I .
\end{align}
It can then be immediately seen that for $p=-(1-k/n)$ and $q=-k/n$ that $I=0$ from (\ref{residue}). We thus have the following result
\begin{align}\label{no-contribution}
 \int_y^{1-y} dz \, \vartheta_1(z-y)^{-(1-k/n)} \vartheta_1(z-1+y)^{-k/n} \vartheta_1 (z - y_M )=0.
\end{align}
Similarly, it can be also be checked that 
\begin{align}
 \int_y^{1-y} d\bz \, \vartheta_1(\bz - y)^{-k/n}\vartheta_1(\bz-1+y)^{-(1-k/n)} \vartheta_1 (\bz + y_M  ) =0 .
\end{align}
So, for $W_1^a$ which involves integration along the real line with this branch cut we conclude that they are real. 

Now, consider the following integral over the spatial cycle
\begin{align}
W_1^1 &= \int _0^1 dz \,  \vartheta_1(z-y)^{-(1-k/n)} \vartheta_1(z-1+y)^{-k/n} \vartheta_1 (z - y_M ) \nn \\
&= \left( \int_0^y + \int_y^{1-y} + \int_{1-y}^1 \right) dz \,  \vartheta_1(z-y)^{-(1-k/n)} \vartheta_1(z-1+y)^{-k/n} \vartheta_1 (z - y_M ).
\end{align}
Note that $z$ is a real variable. The second integral vanishes as we had just shown. Using the double periodicity property $\vartheta_1(z)=\vartheta_1(z+1)$, the third integral can be set from $-y$ to $0$. So, we have
\begin{align}
W_1^1 &=   \left( \int_0^y +  \int_{-y}^0 \right) dz \,  \vartheta_1(z-y)^{-(1-k/n)} \vartheta_1(z-1+y)^{-k/n} \vartheta_1 (z - y_M ).
\end{align}
Similarly,
\begin{align}
W_1^2 &=   \left( \int_0^y +  \int_{-y}^0 \right)  d\bz \, \vartheta_1( \bz - y)^{-k/n}\vartheta_1(\bz-1+ y)^{-(1-k/n)} \vartheta_1 (\bz + y_M ).
\end{align}Again $\bz$ is real. On changing variables in the above integral $\bz = - z$ and using the fact that $\vartheta_1(-z) = -\vartheta_1(z)$ it can be seen that 
\begin{align}
W_1^1 = W_1^2.
\end{align}
We have thus proved the first two relations in (\ref{reality}).

Let us now consider the integrals over the temporal cycle. The integration is now over a purely imaginary variable. We also have $\tau = i\beta$. 
\begin{align}
W_2^1 &= \int _0^\tau dz \,  \vartheta_1(z-y)^{-(1-k/n)} \vartheta_1(z-1+y)^{-k/n} \vartheta_1 (z - y_M ) .
\end{align}
Changing variables $z=iw$ ($w$ is real), we have
\begin{align}
W_2^1 &= \int _0^\beta i dw \,  \vartheta_1(iw-y)^{-(1-k/n)} \vartheta_1(iw-1+y)^{-k/n} \vartheta_1 (iw - y_M ) .
\end{align}
On complex conjugation this becomes
\begin{align}
(W_2^1)^* &= \int _0^\beta -i dw \,  \vartheta_1(-iw-y)^{-(1-k/n)} \vartheta_1(-iw-1+y)^{-k/n} \vartheta_1 (-iw - y_M ) .
\end{align}
On changing variables again $u=-w$ we get
\begin{align}
(W_2^1)^* &= \int _0^{-\beta} i du \,  \vartheta_1(iu-y)^{-(1-k/n)} \vartheta_1(iu-1+y)^{-k/n} \vartheta_1 (iu - y_M ) .
\end{align}
We now use the property, $ \vartheta_1(i(u+\beta)) =  \vartheta_1(iu) $ to get
\begin{align}\label{pure-im}
(W_2^1)^* &= \int _\beta^0 i du \,  \vartheta_1(iu-y)^{-(1-k/n)} \vartheta_1(iu-1+y)^{-k/n} \vartheta_1 (iu - y_M ) \nn \\
&= - W_2^1.
\end{align}
On performing the same analysis it can be shown that $(W_2^2)^* = - W_2^2$. 

Now let's consider the following integral 
\begin{align}
W_2^2 &= \int_0^\tau d\bz \, \vartheta_1( z -  y)^{-k/n}\vartheta_1(\bz-1+  y)^{-(1-k/n)} \vartheta_1 (\bz + y_M ) .
\end{align}
Changing variables $z=iw$ or $\bz = -iw$, with $w$ being real, we get
\begin{align}
W_2^2 &= \int_0^\beta -idw \, \vartheta_1(  -iw -  y)^{-k/n}\vartheta_1( -iw-1+  y)^{-(1-k/n)} \vartheta_1 ( -iw + y_M ) .
\end{align}
Using $\vartheta_1(-z)=-\vartheta_1(z)$, we get 
\begin{align}
W_2^2 &= - \int_0^\beta idw \, \vartheta_1(  iw +  y)^{-k/n}\vartheta_1( iw+1- y)^{-(1-k/n)} \vartheta_1 ( iw -y_M ) \nn \\
&=- W_2^1 = (W_2^1)^*.
\end{align}
The second and third equalities follow from (\ref{pure-im}). This completes the proofs of all relations in  (\ref{reality}).

\subsection*{Relating $W_1^1$ to $W_2^2$}
In this subsection we shall relate $W_1^1$ and $W_2^2$. Choosing the interval as $\left[ \frac{1-L}{2} , \frac{1+L}{2}\right]$ the expressions are
\begin{align}
W_1^1 &= \int_0^1 dz \frac{\vt_1(z-1/2+L/2-(kL)/n)}{\vt_1(z-1/2+L/2)^{1-k/n} \vt_1(z-1/2-L/2)^{k/n}}, \\
W_2^2 &= \int _0^\tau d\bz \frac{\vt_1(\bz-1/2-L/2+(kL)/n)}{\vt_1(\bz-1/2-L/2)^{1-k/n} \vt_1(\bz-1/2+L/2)^{k/n}}.
\end{align}
The integrands of the above are clearly related by $L \rightarrow -L$. Defining 
\begin{align}\label{w-def}
w(z,L) =  \frac{\vt_1(z-1/2-L/2-(kL)/n)}{\vt_1(z-1/2+L/2)^{1-k/n} \vt_1(z-1/2-L/2)^{k/n}} ,
\end{align}
we can write 
\begin{align}\label{new-int}
W_1^1 = \int_0^1 dz \, w(z,L) \ , \qquad W_2^2 = \int_0^\tau d\bz \, w(\bz,-L) .
\end{align}
Since $w(z,L)$ is constructed purely of theta functions, motivated by \eqref{theta-def} let us now consider its Fourier series expansion 
\begin{align}\label{laurent1}
w(z,L) = a_0 (L) + \sum_{m\neq 0} a_m(L) \, e^{2\pi i mz}.
\end{align}
It is assumed that $a_m$ has an implicit $\tau$ dependence.

It can be seen from \eqref{w-def} that the function $w(z,L)$ is invariant under $z \rightarrow z+ \tau$. We thus have the following identity relating the Fourier coefficients.
\begin{align}
\sum_{m\neq 0} a_m(L) \, e^{2\pi i mz} = \sum_{m\neq 0} a_m(L) \, e^{2\pi im (z+\tau)}.
\end{align}
Performing a indefinite integral over $z$ to the above relation, we get
\begin{align}\label{coe-rel}
\sum_{m\neq 0} \frac{a_m(L) \, e^{2\pi i mz}}{2\pi i m } = \sum_{m\neq 0} \frac{a_m(L) \, e^{2\pi i m(z+\tau)}}{2\pi i m}.
\end{align}
The form of the integral for $W_2^2$ is
\begin{align}
W_2^2= \int_0^\tau dz \, w(z,- L) &= \int_0^\tau dz\left[ a_0 (-L) + \sum_{m\neq 0} a_m(-L) \, e^{2\pi i mz} \right] \nn \\
&= \left[  a_0(-L)z +   \sum_{m\neq 0} \frac{a_m(-L) \, e^{2\pi i mz}}{2\pi i m } 	\right]_0^\tau,
\end{align}
which upon using \eqref{coe-rel} is simply
\begin{align}
W_2^2 = a_0 (-L)\tau .
\end{align}
One can however perform an explicit integration of $W_2^2$ order by order in $q(=e^{-2\pi/\beta})$ and then verify that $W_2^2$ is an even function in $L$ (see equation \eqref{hightemp2} below). We therefore conclude that $a_0(L)=a_0(-L)$ and
\begin{align} \label{w2-zero-mode}
W_2^2 = a_0 (L)\tau .
\end{align}
Similarly, the $W_1^1$ integral \eqref{new-int} can be easily done and we get
\begin{align} \label{w1-zero-mode}
W_1^1 = a_0(L) .
\end{align}
Comparing \eqref{w2-zero-mode} and \eqref{w1-zero-mode} we have the following relation
\begin{align} \label{w-rel}
W_2^2 = \tau W_1^1.
\end{align}

The expression for $W_2^2$ can be also  related to the $W_1^1$ in another
way. 
\begin{align}
W_2^2 &= \int_0^\tau d\bz \frac{\vt_1(\bz - 1/2 -L/2 +kL/n)}{\vt_1(\bz
-1/2+L/2)^{k/n} \vt_1 (\bz -1/2-L/2)^{(1-k/n)}}
\end{align}
Performing a S-modular transformation to the above integrand we get
\begin{align}
W_2^2 &= e^{\frac{\pi L^2}{\beta}\frac{k}{n}(1-\frac{k}{n})}  \int_0^\tau
d\bz \frac{\vt_1(\tfrac{\bz - 1/2 -L/2
+kL/n}{\tau}|-\frac{1}{\tau})}{\vt_1(\tfrac{\bz - 1/2 +L/2
}{\tau}|-\frac{1}{\tau})^{k/n} \vt_1(\tfrac{\bz - 1/2
-L/2}{\tau}|-\frac{1}{\tau})^{(1-k/n)}}
\end{align}
Now, changing the integration variable from $\bz$ to $w=-\bz /\tau$ and
then shifting the range of integration from $[-1,0]$ to $[0,1]$ (which is
allowed by periodicity of the integrand) we get
\begin{align}
W_2^2 &= \tau e^{\frac{\pi L^2}{\beta}\frac{k}{n}(1-\frac{k}{n})}  \int_0^1
dw \frac{\vt_1(w+\tfrac{  1/2 +L/2
-kL/n}{\tau}|-\frac{1}{\tau})}{\vt_1(w+\tfrac{1/2 -L/2
}{\tau}|-\frac{1}{\tau})^{k/n} \vt_1(w+\tfrac{ 1/2
+L/2}{\tau}|-\frac{1}{\tau})^{(1-k/n)}}
\end{align}
We can now use the periodicity properties of $\vt_1(z/\tau|-1/\tau)$ and
shift the argument  by $-1/\tau$ in the above integrand to get
\begin{align}
W_2^2 &= \tau e^{\frac{\pi L^2}{\beta}\frac{k}{n}(1-\frac{k}{n})}  \int_0^1
dw \frac{\vt_1(w+\tfrac{  -1/2 +L/2
-kL/n}{\tau}|-\frac{1}{\tau})}{\vt_1(w+\tfrac{-1/2 -L/2
}{\tau}|-\frac{1}{\tau})^{k/n} \vt_1(w+\tfrac{ -1/2
+L/2}{\tau}|-\frac{1}{\tau})^{(1-k/n)}}
\end{align}
The integral appearing above is the that of $W_1^1$ with the replacements :
$L \rightarrow L/\tau$   and $\tau = -1/\tau$. We finally have the following result\footnote{In the arguments of $\vt_1$ in the integrand of $W_1^1$ 1/2 gets replaced by $1/(2\tau)$. This can be thought of as the length of the spatial circle $R$ being replaced by $R/\tau$.}
\begin{align}
W_2^2 (L, \tau) = \tau e^{\frac{\pi L^2}{\beta}\frac{k}{n}(1-\frac{k}{n})}
W_1^1 (L/\tau,   -1/\tau)
\end{align}

\subsection*{High temperature expansions of $W_1^1$ and $W_2^2$}
We have
\begin{align}
W_2^2 &= \int_0^\tau d\bz \frac{\vartheta_1(\bz - \frac{1}{2} - \frac{L}{2} + \frac{kL}{n}  )}{\vartheta_1(\bz - \tfrac{1}{2} + \tfrac{L}{2})^{k/n}\vartheta_1(\bz - \tfrac{1}{2} - \tfrac{L}{2})^{1-k/n}}.
\end{align}
Using the periodicity property of the $\vartheta_1(z|\tau)$ 
\begin{align}
\vartheta_2(z|\tau) = - \vartheta_1 (z-1/2 |\tau) ,
\end{align}
we get
\begin{align}
W_2^2 &= \int_0^\tau d\bz \frac{\vartheta_2(\bz   - \frac{L}{2} + \frac{kL}{n}  )}{\vartheta_2(\bz   + \tfrac{L}{2})^{k/n}\vartheta_2(\bz  - \tfrac{L}{2})^{1-k/n}}.
\end{align}
Performing a S-modular transformation to the above integrand
\begin{align}
W_2^2 = e^{\frac{\pi}{\beta} \frac{k}{n} (1-\frac{k}{n})L^2} \int_0^\tau d\bz \frac{  \vartheta_4(\frac{\bz -L/2 +kL/n}{\tau}|-\frac{1}{\tau})   }{\vartheta_4(\frac{\bz +L/2 }{\tau}|-\frac{1}{\tau})^{k/n}  \vartheta_4(\frac{\bz -L/2 }{\tau}|-\frac{1}{\tau})^{1-k/n}}.
\end{align}
Changing variables to $w$, where $z=iw$ and $w \in \mathbb{R}$ since the integral is along  the negative imaginary axis 
\begin{align}
W_2^2 = -ie^{\frac{\pi}{\beta} \frac{k}{n} (1-\frac{k}{n})L^2} \int_0^{-\beta} dw \frac{  \vartheta_4(\frac{-iw -L/2 +kL/n}{\tau}|-\frac{1}{\tau})   }{\vartheta_4(\frac{-iw +L/2 }{\tau}|-\frac{1}{\tau})^{k/n}  \vartheta_4(\frac{-iw -L/2 }{\tau}|-\frac{1}{\tau})^{1-k/n}}.
\end{align}
We can now expand the above integral in powers of $q=e^{-2\pi/ \beta}$, 
\begin{align}\label{hightemp2}
&W_2^2 (k,n)\nn \\= & \; e^{\frac{\pi}{\beta} \frac{k}{n} (1-\frac{k}{n})L^2}\tau \nn \\ & \times \Big{[} 1  -\frac{2}{n^2} {\left(-k^2+k n \cosh \left(\tfrac{2 \pi  L (n-k)}{n\beta}\right)+(k-n) \left(k \cosh \left(\tfrac{2 \pi  L}{\beta}\right)-n \cosh \left(\tfrac{2 \pi  k L}{n\beta }\right)\right)+k n-n^2\right)} q^2\nn \\
&\quad \ \ + \frac{1}{2 n^4}\left(2 n \left(3 (k-n) \left(k^2-k n+2 n^2\right) \cosh \left(\tfrac{2 \pi  k L}{n}\right) \right. \right. \nn \\ 
&\qquad\qquad \qquad \left. \left. +k \left(-3 \left(k^2-k n+2 n^2\right) \cosh \left(\tfrac{2 \pi  L (n-k)}{n}\right)-(k-2 n) (k-n) \cosh \left(\tfrac{2 \pi  L (k+n)}{n}\right) \right. \right.\right. \nn \\
&\qquad\qquad \qquad \qquad\qquad\left. \left. \left. +(k-n) (k+n) \cosh\left(\tfrac{2 \pi  L (k-2 n)}{n}\right)\right)\right) \right. \nn \\ 
&\qquad\qquad \left.-4 k (k-n) \left(k^2-k n+4 n^2\right) \cosh (\tfrac{2 \pi  L}{\beta})+3 \left(k^4-2 k^3 n+7 k^2 n^2-6 k n^3+4 n^4\right) \right. \nn \\ 
&\qquad\qquad +k (k-n) (k+n) (k-2 n) \cosh (\tfrac{4 \pi  L}{\beta})\Big{)} q^4 +\mathcal{O}(q^6) \Big{]}.
\end{align}
As remarked earlier it can be clearly seen that the above expression is a even function in $L$. 
The high-temperature expansion of $W_1^1$ can be obtained by using the relation \eqref{w-rel}. 

\subsection*{$W_1^1$ and $W_2^2$ in the $y \rightarrow 0$ limit}

Using the results of the previous subsections above we shall exactly evaluate the integrals $W_1^1$ and $W_2^2$ in the large interval or $y \rightarrow 0$ limit. For $y=\epsilon$ we have
\begin{align}\label{W-y}
W_1^1 &= \int _0^1 dz \,  \vartheta_1(z-\epsilon)^{-(1-k/n)} \vartheta_1(z-1+\epsilon)^{-k/n} \vartheta_1 (z - y_M ) ,
\end{align}
here, $y_M = (1-\frac{2k}{n})\epsilon + \frac{k}{n}$. The above integral can be written as 
\begin{align} 
W_1^1 &=\left( \int _0^\epsilon + \int_{\epsilon}^{1-\epsilon} +\int_{1-\epsilon}^1 \right) dz \,  \vartheta_1(z-\epsilon)^{-(1-k/n)} \vartheta_1(z-1+\epsilon)^{-k/n} \vartheta_1 (z - y_M ) .
\end{align}
The branch cut here is in the interval $\left[\epsilon,1-\epsilon\right]$. It was seen in \eqref{no-contribution} that there is no contribution to the integral from this region. We thus have
\begin{eqnarray} \label{w1-y0}
W_1^1 &=& \left( \int _0^\epsilon +\int_{1-\epsilon}^1 \right) dz \,  \vartheta_1(z-\epsilon)^{-(1-k/n)} \vartheta_1(z-1+\epsilon)^{-k/n} \vartheta_1 (z - y_M ) \nn \\
 &=&  2 \int _0^\epsilon  dz \,  \frac{ \vartheta_1 (z - y_M )}{\vartheta_1(z-\epsilon)^{(1-k/n)} \vartheta_1(z-1+\epsilon)^{k/n} } \nn \\
 &\simeq&   \lim_{\epsilon\rightarrow 0} 2\epsilon \left( 
 \frac{1}{2} \frac{\vt_1(-k/n)}{\vt_1 (\epsilon )} \right)= 
  \frac{\vt_1(-k/n)}{\vt'_1 (0)}.
\end{eqnarray}
Note that here we have used the fact that integrand is real in the interval 
$[0, \epsilon]$ and $[1-\epsilon , 1]$. 
We have multiplied the integral by $1/2$ to take into  account the 
contribution per unit cell. 
Using \eqref{w-rel} we have
\begin{align}\label{w2-y0}
W_2^2 & \simeq   \tau \frac{\vt_1(-k/n)}{\vt'_1 (0)}.
\end{align} 

\def\bq{\bar{q}}
\section{One-loop determinant for the Chern-Simons gauge field}
\label{chern-simons}
In this subsection we shall calculate the one-loop determinant for the Chern-Simons spin-1 field. 
We present this calculation in detail since we did not find an explicit discussion of this
in the literature. 
The one-loop determinant is given in terms of the product of the one-loop determinants of the scalar and the transverse vector Laplacian \cite{Gegenberg:1993gd}.
\begin{align}
Z_\ol = \det\null^{1/2} (-\Delta_{(0)} ) \det\null ^{-1/4} (- \Delta_{A^\perp} -2) .
\end{align}
The scalar and $A^\perp$ one-loop determinants are as follows
\begin{align}
\frac{1}{2} \log \det (- \Delta_{(0)} ) &= - \sum_{m=1}^\infty \frac{|q|^{2m}}{m|1-q^m|^2} ,\nn \\
-\frac{1}{4} \log \det ( - \Delta_{A^\perp} -2 ) &= \frac{1}{2} \sum_{m=1}^\infty \frac{q^m+\bq^m }{m|1-q^m|}.
\end{align}
Adding the contributions from the two terms above we get
\begin{align}
\log Z_\ol &= \sum_{m=1}^\infty \frac{1}{2m|1-q^m|^2}(-2 (q \bq)^m + q^m +\bq ^m )\nn \\
&=  \sum_{m=1}^\infty \frac{1}{2m|1-q^m|^2} ( 2|1-q^m|^2 - (1-q^m) - (1-\bq ^m))\nn \\
&= - \sum_{m=1}^\infty \frac{1}{m} + \frac{1}{2} \sum_{m=1}^\infty \frac{1}{m(1-q^m)} + \frac{1}{2} \sum_{m=1}^\infty \frac{1}{m(1-\bq^m)} \nn \\
&= \frac{1}{2} \sum_{n=1}^\infty \sum_{n=1}^\infty \frac{q^{nm}}{m} + \frac{1}{2} \sum_{n=1}^\infty \sum_{n=1}^\infty \frac{\bq^{nm}}{m} \nn \\
&= - \frac{1}{2} \sum_{n=1}^\infty \log |1-q^n|^2 .
\end{align}
So, the one-loop determinant of a single $U(1)$  Chern-Simons gauge field is
\begin{align}
Z_\ol = \prod_{n=1}^\infty \left(  \frac{1}{|1-q^n|^2}  \right)^{1/2} .
\end{align}

\section{Sums involving 2-letter words}\label{app-sum}

We need to do the sums over $k_1$ and $k_2$ over the reciprocals of the eigenvalues of the 2-letter words given below.
\begin{align}
q_{\pm \pm}^{-1/2} &= \frac{n^2 }{(u_y^{-1} - u_y)^2 u_R} \left( u_y ^{-1/n} - e^{2\pi i (k_1 -k_2)/n} u_y^{1/n} \right)   \left( u_y ^{-1/n} - e^{-2\pi i (k_1 -k_2)/n} u_y^{1/n} \right) +\mathcal{O}(1) \label{qpp}\\
q_{\pm \mp}^{-1/2} &= -\frac{n^2  }{(u_y^{-1} - u_y)^2 u_R}  \left(1 - e^{2\pi i (k_1 -k_2)/n}   \right)   \left( 1 - e^{-2\pi i (k_1 -k_2)/n}   \right) + \mathcal{O}(1) \label{qmm}
\end{align}
The first sum is of the following form
\begin{align}\label{sum1}
\sum_{k_1=0}^{n-1}{\sideset{}{'}\sum_{k_2=0}^{n-1}} q_{\pm \pm}^{1/2} &= \frac{4 \sinh ^2 (2\pi T y) e^{-2\pi T R}}{n^2 e^{4\pi T y / n }} \sum_{k_1 =0}^{n-1} {\sideset{}{'}\sum_{k_2 =0}^{n-1}} \frac{1}{(1-r e^{i(k_1-k_2 )2\pi /n})(1-r e^{-i(k_1-k_2 )2\pi /n})}
\end{align}
where $r=u_y^{2/n}$. Let us now do the sum above. Implementing the $k_1\neq k_2$ constraint by removing the `diagonal' $k_1 = k_2$ piece and counting $(k_1,k_2)$ pairs only once we get
\begin{align}
&\frac{1}{2}\left( \sum_{k_1 =0}^{n-1} \sum_{k_2 =0}^{n-1}  \frac{1}{(1-r e^{i(k_1-k_2 )2\pi /n})(1-r e^{-i(k_1-k_2 )2\pi /n})}       - \frac{n}{(1-r)^2}  \right)  \nn \\
=& \frac{1}{2}\left( \sum_{k_1 =0}^{n-1} \sum_{k_2 =0}^{n-1}   \sum_{l=0}^\infty r^l e^{2\pi i l (k_1-k_2)/n}   \sum_{s=0}^\infty r^s e^{-2\pi i s (k_1-k_2 )/n}    - \frac{n }{(1-r)^2}\right)  \\
=&\frac{1}{2}\left(  \sum_{l=0}^\infty   \sum_{s=0}^\infty    r^{l+s}  \sum_{k_1 =0}^{n-1}  e^{2\pi i k_1(l-s )/n}  \sum_{k_2 =0}^{n-1}  e^{-2\pi i k_2 (l-s )/n}   - \frac{n}{(1-r)^2} \right)   \nn 
\end{align}
The sums over $k_{1,2}$ are non-zero only if $l-s=Mn$ where $M \in \mathcal{Z}$. So, one can write the above sum as
\begin{align}
\frac{1}{2}\left(  \sum_{l=0}^\infty   \sum_{s=0}^\infty    r^{l+s}  \sum_{M=0}^{\infty} (n\delta_{l-s,M})   - \frac{n}{(1-r)^2} \right)  
\end{align}
where
\begin{align}
&l= Mn + s \ , \qquad s= Nn+j \nn \\
&s \in \lbrace 0,1,2, \cdots \rbrace \quad j \in \lbrace 0,1,2, \cdots , n-1\rbrace \quad n \in \lbrace 0,1,2, \cdots \rbrace 
\end{align}
Also, $l>0$ implies $Mn + Nn +j >0$, which means $M$ can take the lowest value $-N$.
Converting the above sum over $l$, $s$ and $M$ to $N$, $M$ and $j$, we get
\begin{align}
&\frac{1}{2}\left( n^2\sum_{N=0}^\infty \sum_{j=0}^{n-1} \sum_{M=-N}^\infty r^{Mn + 2 (Nn+j)} -n \sum_{l=0}^\infty   \sum_{s=0}^\infty    r^{l+s}\right)\nn \\
=&\frac{1}{2}\left(n^2\sum_{N=0}^\infty \sum_{j=0}^{n-1}  r^{2j + 2 Nn} \sum_{M=-N}^\infty r^{Mn} - \frac{n}{(1-r)^2} \right)\nn \\
=& \frac{1}{2}\left(n^2\sum_{N=0}^\infty \sum_{j=0}^{n-1}  r^{2j + 2 Nn}  \frac{r^{-Nn}}{1-r^n}  - \frac{n}{(1-r)^2} \right)\nn \\
=& \frac{1}{2}\left(n^2\frac{1}{(1-r^n)^2} \frac{1-r^{2n}}{1-r^2}  - \frac{n}{(1-r)^2} \right) \nn \\
=& \frac{1}{2}\left(n^2\frac{r^{-1}}{r^{-1}-r} \; \frac{r^{-n/2} + r^{n/2}}{r^{-n/2}- r^{n/2}}  - \frac{n}{(1-r)^2}  \right)\nn \\
=& \frac{1}{2}\left(n^2\frac{e^{4\pi T y/n}}{2 \sinh (\frac{4\pi Ty}{n})} \coth (2\pi T y)  - n \frac{e^{4\pi Ty/n}}{4\sinh^2 (2\pi Ty/n)}                  \right)\label{r-sum}
\end{align}
Using this in (\ref{sum1}) we get
\begin{align} \label{sum111}
\sum_{k_1=0}^\infty {\sideset{}{'}\sum_{k_2=0}^\infty} q_{\pm\pm}^{1/2}  &=  \frac{\sinh (\tfrac{4\pi T y}{n})}{2\sinh ( {4\pi T y} )} e^{-2\pi TR} - \frac{\sinh^2 (\tfrac{2\pi T y}{n})}{2n\sinh^2 ( {2\pi T y} )} e^{-2\pi TR} + \mathcal{O}(e^{-4\pi TR}) 
\end{align}
The second sum \eqref{qmm} is 
\begin{align}\label{sum2}
\sum_{k_1=0}^\infty{\sideset{}{'}\sum_{k_2=0}^\infty} q_{\pm \mp}^{1/2} &= -\frac{4 \sinh ^2 (2\pi T y) e^{-2\pi T R}}{n^2  } \nn \\ & \qquad \times \lim_{r\rightarrow 1^-} \sum_{k_1 =0}^{n-1}{\sideset{}{'}\sum_{k_2=0}^\infty}\frac{1}{(1-r e^{i(k_1-k_2 )2\pi /n})(1-r e^{-i(k_1-k_2 )2\pi /n})}
\end{align}
\def\nm{{n-1}}
Let us now do the sum above. Taking into account the constraints which were present while doing the previous sum we have
\begin{align}\label{2ndsum}
&\frac{1}{2}\left( \sum_{k_1=0}^{\nm} \sideset{}{}\sum_{k_2=0}^{\nm} \frac{1}{(1-r e^{i(k_1-k_2 )2\pi /n})(1-r e^{-i(k_1-k_2 )2\pi /n})} - \frac{n}{(1-r)^2} \right) \nn \\
=&\frac{1}{2}\left( \sum_{k_1=0}^{\nm} \sideset{}{}\sum_{k_2=0}^{\nm} \sum_{l=0}^\infty r^l e^{2\pi i l (k_1-k_2)/n}   \sum_{s=0}^\infty r^s e^{-2\pi i s (k_1-k_2 )/n} - \frac{n}{(1-r)^2} \right)\nn \\
=& \frac{1}{2}\left(\sum_{l=0}^\infty   \sum_{s=0}^\infty    r^{l+s} \sum_{k_1=0}^{\nm} e^{2\pi i k_1(l-s )/n}   \sideset{}{}\sum_{k_2=0}^{\nm}  e^{-2\pi i k_2 (l-s )/n} - \frac{n}{(1-r)^2} \right)\nn \\
=&\frac{1}{2}\left( \sum_{l=0}^\infty   \sum_{s=0}^\infty    r^{l+s}  \sum_{k_1=0}^{\nm} e^{2\pi i k_1(l-s )/n}  \sum_{k_2=0}^{\nm}  e^{-2\pi i k_2 (l-s )/n} -  \frac{n}{(1-r)^2} \right)
\end{align}
The first term in \eqref{2ndsum} is the  same as the one calculated in the previous sum \eqref{r-sum}. So, the above sum becomes
\begin{align}
\frac{1}{2} \left(  n^2\frac{1}{(1-r^n)^2} \frac{1-r^{2n}}{1-r^2} - n \frac{1}{(1-r)^2}  \right)
\end{align}
One can now take the limit $r\rightarrow 1^-$. We get
\begin{align}
\lim_{r\rightarrow 1^-} \frac{1}{2n^2} \left[ n^2\frac{1}{(1-r^n)^2} \frac{1-r^{2n}}{1-r^2} - n \frac{1}{(1-r)^2} \right] =
\frac{1}{24n}   \left(n^2-1\right)
\end{align}
Using the above in (\ref{sum2}) we obtain
\begin{align}\label{sum222}
\sum_{k_1=0}^\infty {\sideset{}{'}\sum_{k_2=0}^\infty} q_{\pm\mp}^{1/2} &= -\frac{ (n^2 -1)}{6n} \sinh^2 (2\pi T y) e^{-2\pi TR}  + \mathcal{O}(e^{-4\pi TR}) 
\end{align}
The above sums \eqref{sum111} and \eqref{sum222} are used in \eqref{fermion-sum} in order to obtain \eqref{fermion-Z}.

\section{Details of Mathematica files}
\label{mathematica}
In the main text we have shown the plots of the \Re entropies at different
temperatures and have numerically extrapolated the \Re entropy to
$n\rightarrow 1$ to obtain the entanglement entropy. The relevant
Mathematica files are attached with the arXiv source.
\begin{enumerate}
\item   \underline{renyi.nb} performs the numerical integration of $W_2^2
(k,n)$. The integration is done using the default adaptive quadrature
technique in  Mathematica. The answer for $W_2^2$ is then used to calculate
the \Re entropy from \eqref{renyi-1}. We then plot the $n=2$ \Re entropies
at different temperatures and the $n=2,3,4$ and $5$ \Re entropies at
$\beta=0.6$ (Fig\,\ref{renyi-plot}).
\item \underline{extrapolation.nb} contains the computation for numerical
extrapolation of $n=2,3,\cdots,10$ \Re entropies to $n=1$ in order to obtain the entanglement entropy. This is done in
Mathematica by fitting 3rd order polynomial curves (Fig\,\ref{extra-plot}).
\item \underline{high-temp.nb} is a program to analytically evaluate the
high temperature expansion of \Re and entanglement entropies of free bosons
on  the torus upto any order in $e^{-2\pi TR}$. The expression for the
entanglement entropy calculated using this upto $\mathcal{O}(e^{-6\pi TR})$
is given in equation \eqref{q4}.
\end{enumerate}

\bibliography{bosons}
\bibliographystyle{JHEP}
\end{document}